\documentclass[twocolumn, trackchanges]{aastex701}
\usepackage{amsmath}
\usepackage{newtxtext,newtxmath}
\usepackage{xcolor}
\usepackage{adjustbox}
\usepackage{graphicx}
\usepackage{amsmath}
\usepackage{footnote}
\usepackage{amssymb}
\usepackage{multirow}

\def \inte {{\it INTEGRAL}}

\def \xmm {{\it XMM}-Newton}
\def \src{{SRGA~J144459.2-604207}}

\def \nustar{{\it NuSTAR}}

\def \nicer{{\it NICER}}

\def \ixpe{{\it IXPE}}

\def \ninjasat{{\it NinjaSat}}

\shorttitle{Burst oscillation in \src{}}
\shortauthors{Mandal \& Naik}
\begin{document}

\title{Discovery of burst oscillations in the newly discovered millisecond X-ray pulsar \src{}}

\correspondingauthor{Manoj Mandal}
\email{manojmandal213@gmail.com}
\author[orcid=0000-0002-1894-9084]{Manoj Mandal}
\affiliation{Astronomy and Astrophysics Division, Physical Research Laboratory, Navrangpura, Ahmedabad - 380009, Gujarat, India}
\email{manojmandal213@gmail.com}

\author[orcid=0000-0003-2865-4666]{Sachindra Naik} 
\affiliation{Astronomy and Astrophysics Division, Physical Research Laboratory, Navrangpura, Ahmedabad - 380009, Gujarat, India}
\email{snaik@prl.res.in}

%\accepted{for publication in The Astrophysical Journal, DOI : XX}

\begin{abstract}
 Burst oscillations during thermonuclear X-ray bursts are powered by thermonuclear energy on the neutron star (NS) surface and typically occur close to the spin frequency of the NS. We performed a comprehensive timing analysis of all thermonuclear bursts from the newly discovered millisecond X-ray pulsar \src, observed with \nicer, \xmm, and \nustar{} during the 2024 outburst. A total of 39 bursts were detected, allowing for a detailed search for burst oscillations, which had not been previously observed from this source. We report the discovery of burst oscillations at 447.7-448.0 Hz from \src{} using \xmm{} and \nustar{} data, consistent with the spin frequency of the NS. The strongest burst oscillation in the \xmm~ data occurred with a single-trial significance of 5.1$\sigma$ and maximum $Z^{2}$ power of $\sim$31. In the \nustar~ data, the strongest oscillation signal has a significance of 5.2$\sigma$ and maximum $Z^{2}$ power of $\sim$32. The folded pulse profile corresponding to the strongest signal in the 0.5–10 keV band of the \xmm{} data shows a sinusoidal shape with a fractional rms amplitude of $\sim$8.5\%, while the measurements of the \nustar{} data (3-40 keV range) yield $\sim$21\%. These results represent the first detection of burst oscillations in \src. Additionally, we report detection of 447.6~Hz oscillations occurring just before a burst onset observed with \xmm. This marks only the second instance in which burst oscillations have been observed before the burst onset.
\end{abstract}


\section{Introduction}
\label{intro}
A neutron star (NS) low-mass X-ray binary (LMXB) consists of a neutron star and a low-mass companion star ($M_d \leq 1,M_\odot$) orbiting the common center of mass. In these systems, matter from the companion is transferred to the NS via the Roche-lobe overflow mechanism, forming an accretion disk. In high magnetic field NS LMXBs ($\sim10^{12}$ G; \citealt{Staubert2019}), the disk is truncated at a relatively large magnetospheric radius. In contrast, the low magnetic field ($\sim10^{7-9}$ G; \citealt{Ca09, Mu15}) of the NS in the LMXBs allows the disk to extend close to the NS surface. In this case, the accreted materials accumulate directly on the NS surface rather than being funneled to the magnetic poles of the NS. The accumulated hydrogen and helium can undergo unstable nuclear burning, producing thermonuclear (Type-I) X-ray bursts \citep{Le93, Ga06}. During bursts, the X-ray intensity of the source increases rapidly, often exceeding persistent emission by more than an order of magnitude \citep{Ba10}. Bursts are typically of short duration, with fast rise times (0.5–5 s) and slow exponential decays (10–100 s) \citep{Le93}. During the burst, lighter elements fuse into heavier nuclei through nuclear chain reactions \citep{Le93, St03, Sc06}.

During a few such bursts in several LMXBs, burst oscillations are observed at a frequency typically in the range of a few tens to several hundred hertz, close to the NS spin frequency \citep{Watts2012, Ga08}. These oscillations are believed to originate from asymmetric brightness patterns on the surface of the NS during unstable nuclear burning. The detection of burst oscillations provides a powerful tool to constrain the fundamental neutron star parameters, including spin, mass, and radius \citep{Watts2012}, and also provides important insights into the physics of ultra-dense matter in high magnetic field environments \citep{Strohmayer2006, Watts2012}. In addition to burst oscillations, quasi-periodic oscillations (QPOs) are often detected during X-ray bursts, allowing for further probing of the dynamics of the accretion flow and the neutron star surface \citep{Homan1999, Wang2025}.

Accreting millisecond pulsars (AMXPs) are a subclass of NS LMXBs that exhibit coherent millisecond pulsations (spin frequencies in the 180-600 Hz range; \citealt{Po06, Salvo2020, Alessandro2021}) and thermonuclear X-ray bursts \citep{Bhattacharyya2022}. These systems provide a unique laboratory to study unstable burning and its interaction with the truncated accretion disk. The accreting millisecond pulsar \src{} was discovered in February 2024 \citep{Me24} with the Mikhail Pavlinsky ART-XC telescope onboard the Spectrum–Roentgen–Gamma (SRG) observatory \citep{Pa21, Su21} during an X-ray outburst that lasted for nearly a month. Shortly after its discovery, the source was targeted by an extensive multi-wavelength follow-up campaign. \citet{Ng24} discovered coherent X-ray pulsation at $\sim$447.9 Hz in the AMXP \src~ and also determined the orbital period of the binary as $\sim$5.2 hours using \nicer~ data. The \nicer{} observation also revealed a thermonuclear X-ray burst from the source \citep{Ng24}. Subsequently, additional thermonuclear bursts are observed with several high-energy missions, including \inte{} \citep{Sa24a, Sa24b}, \ninjasat{} \citep{Ta24, Takeda2025}, \xmm{}, and \nustar{} \citep{Malacaria2025, SRGA25}. The \ixpe{} observation revealed polarized X-ray emission from \src{}, measuring an average polarization degree of $\sim$2.3\% \citep{Pa25}, providing new insight into the geometry of the system and its emission processes. 

In this work, we present the results from a comprehensive timing analysis of all thermonuclear bursts detected from the AMXP \src{} with \nicer, \nustar, and \xmm~ during the 2024 observation campaign. A total of 39 thermonuclear X-ray bursts are detected in the \nicer, \xmm~ EPIC-PN, and \nustar~ datasets, among which nine are simultaneously observed with \nustar~ and \xmm. The broad, high-quality coverage of these observations provides an excellent opportunity to search for burst oscillations, which have not yet been detected from this source. Here, we report the discovery of burst oscillations from \src{} using observations from both \xmm~ and \nustar. We present results from a detailed timing analysis, including pulse-profile modeling and measurements of the fractional rms amplitude. The paper is organized as follows: Section~\ref{obs} describes the observations and data reduction; Section~\ref{res} presents results from our timing analysis, and Section~\ref{dis} discusses the implications and summarizes the conclusions.

%%%%%%%%%%%%%%%%%%%%%%%%%%%%%%%%%%%%%%%%%%%%%%%%%%%%%%%%%%%%%%%%%%%%%%%%%%%%%
  \begin{figure}
\centering{
\includegraphics[width=8.5cm, angle=0]{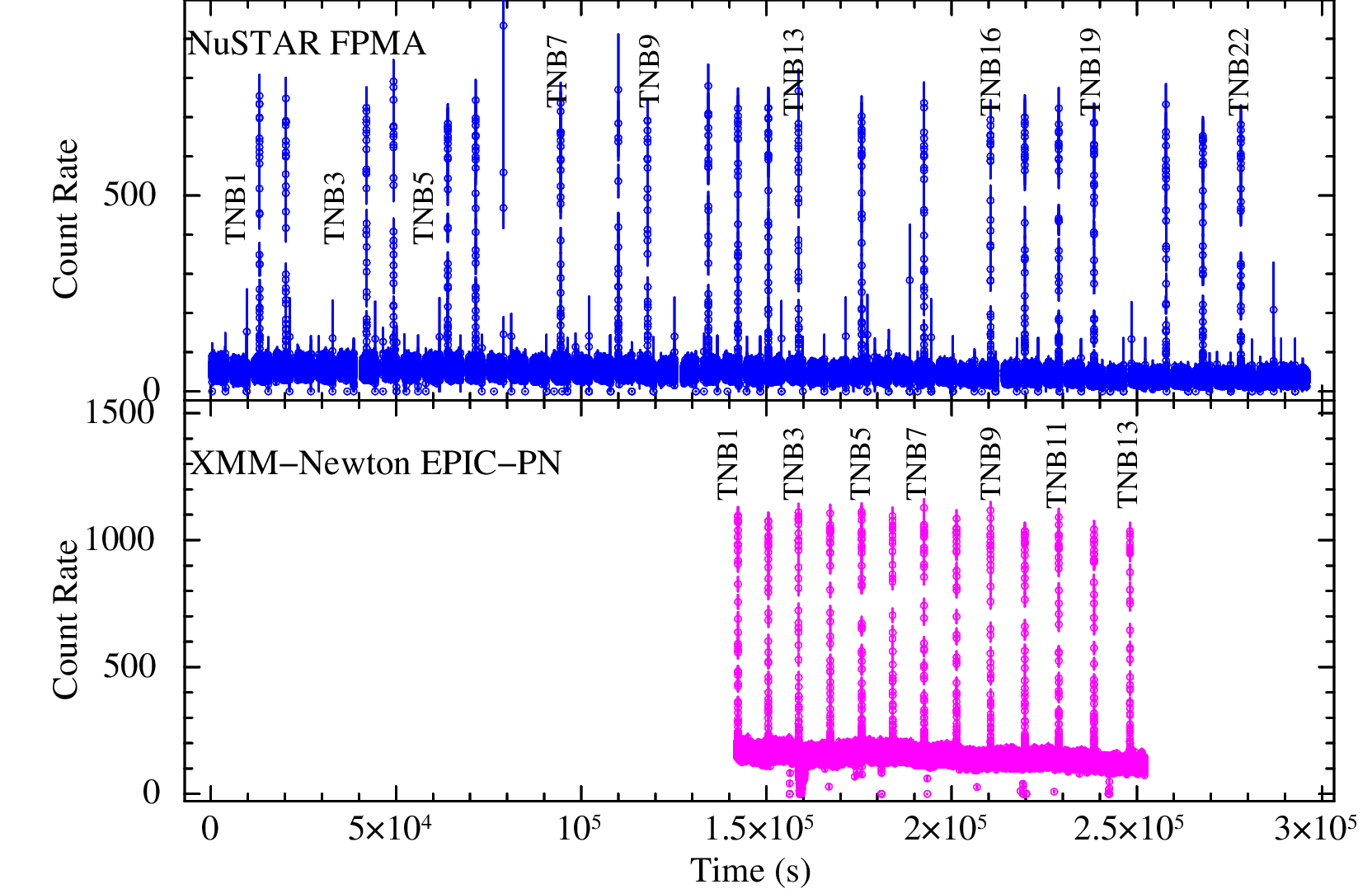}
\caption{{The 1-s binned light curves of \src{} from the \nustar{} and \xmm{} observations, showing several thermonuclear X-ray bursts. Each thermonuclear bursts are numbered as given in both panels.} 
}
\label{fig:burst_lc}}
\end{figure} 
%%%%%%%%%%%%%%%%%%%%%%%%%%%%%%%%%%%%%%%%%%%%%%%%%%%%%%%%%%%%%%%%%%%%%%%%%%

\section{Observation and data analysis}
\label{obs}
In this work, we used publicly available data from the \xmm, \nicer, and \nustar~ observations of \src{} in February 2024. 
 %%%%%%%%%%%%%%%%%%%%%%%%%%%%%%%%%%%%%%%%%%%%%%%%%%%%%5
\subsection{\xmm{} observation}
The AMXP \src{} was observed with \xmm~ on 27 February 2024, for a total exposure of $\sim$135 ks (Obs. ID - 0923171501, PI: A. Papitto). We used European Photon Imaging Camera (EPIC) PN timing-mode data in this study. The \xmm{} Science Analysis System ({\tt SAS}) version 20.0.0 is used to analyze the \xmm~ data. The EPIC-PN event files were produced with {\tt epproc} and filtered using {\tt evselect} with PATTERN $\le$ 4 and FLAG = 0. Source events were extracted from a 20-pixel strip centered on RAWX = 37 (RAWX = 27-47), and background events from a source-free region (RAWX = 5-15). Final source and background light curves were generated with {\tt evselect}, and background subtraction and instrumental corrections were applied using {\tt epiclccorr}. The EPIC-PN timing mode data were barycenter corrected using {\tt barycen} with the JPL DE405 Solar System ephemeris, and source coordinates from \citet{Illiano2024}.
%%%%%%%%%%%%%%%%%%%%%%%%%%%%%%%%%%%%%%%%%%%%%%%%%%%%%%%%%%%%%%%%%%%%%%%%%5
\subsection{\nicer{} observation}
The Neutron Star Interior Composition Explorer (\nicer), a non-imaging soft X-ray telescope operating in the 0.2–12 keV range aboard the International Space Station (ISS) \citep{Ge16}. \nicer~ monitored the AMXP \src{} shortly after its discovery in February 2024. The source was observed during 21-23 February 2024 for a total exposure of $\sim$9 ks through the ToO/DDT program, followed by additional observations under the \nicer~ Guest Observer program PI: A. Papitto). Data reduction was carried out using {\tt HEASOFT} 6.33.2. \nicer~ raw data were processed with {\tt NICERDAS}, and clean, calibrated event files were produced with {\tt nicerl2} using {\tt CALDB} xti20240206. We applied barycentric corrections with {\tt barycorr} in the ICRS frame, using the source coordinates from \citet{Illiano2024} and the JPL DE430 Solar System ephemeris. The light curves were extracted with {\tt XSELECT}, and the results were subsequently used for a detailed timing analysis. During \nicer~ observations, a total of four thermonuclear X-ray bursts were detected (see \citep{Ng24, SRGA25}), none of which are simultaneous with either \xmm~ or \nustar~ observation.
%%%%%%%%%%%%%%%%%%%%%%%%%%%%%%%%%%%%%%%%%%%%%%%%%%%%%
\subsection{\nustar{} observation}
The Nuclear Spectroscopic Telescope Array (\nustar) is an imaging X-ray observatory consisting of two co-aligned focal plane modules, FPMA and FPMB, operating in the 3–79 keV range \citep{Ha13}. The AMXP \src{} was observed on 26 February 2024 for an exposure of $\sim$157 ks (Obs. ID - 80901307002). Data were processed using {\tt NUSTARDAS} in {\tt HEASOFT} 6.33.2 with {\tt CALDB} 20250428. The event files were processed using {\tt nupipeline}. Source events were extracted from a 120 arcsec circular region centered on the source, while a similar off-source region was used for the background. The light curves of source and background for FPMA and FPMB were generated with {\tt nuproducts}. The data were barycenter corrected using {\tt barycorr} with the JPL DE430 Solar System ephemeris and source coordinates from \citet{Illiano2024}. The final background-corrected light curves were produced using {\tt lcmath}. 
%%%%%%%%%%%%%%%%%%%%%%%%%%%%%%%%%%%%%%%%%%%%%%%%%%%%%5
\begin{figure*}
    \adjustbox{valign=t} {\includegraphics[width=0.75\columnwidth]{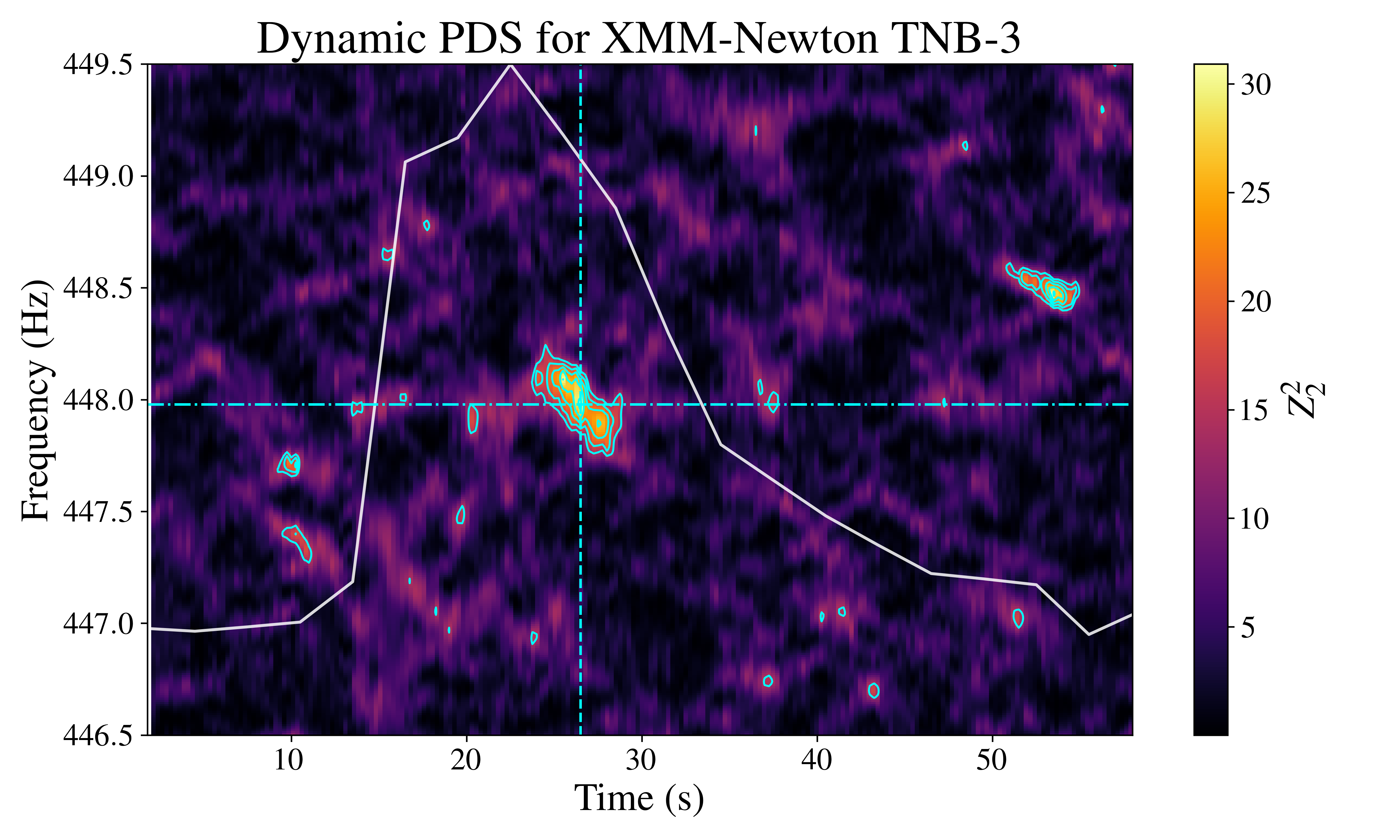}}
  \hspace{-0.03\textwidth}
   \adjustbox{valign=t}{\includegraphics[width=0.65\columnwidth]{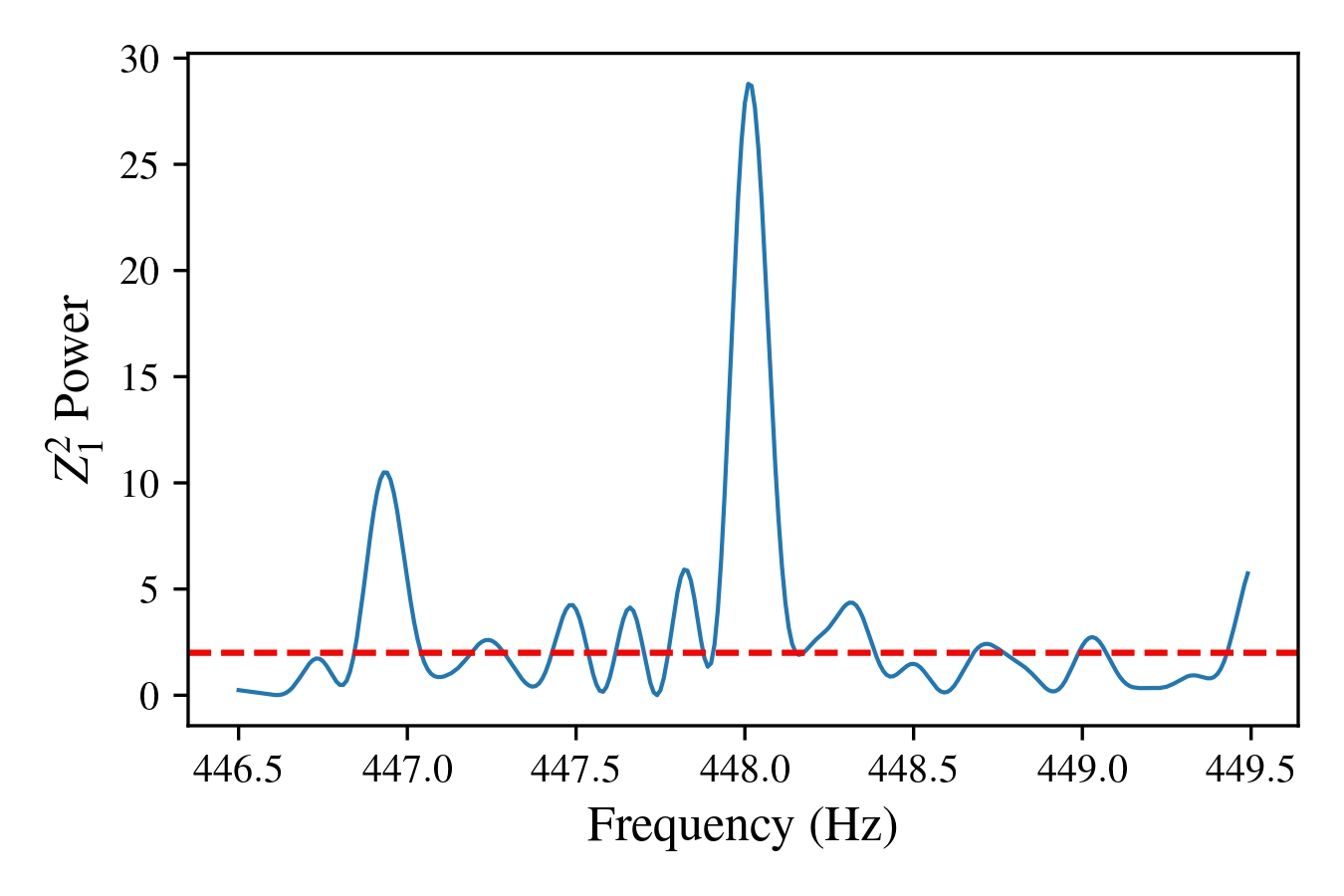}}
    \hspace{-0.02\textwidth}
 \adjustbox{valign=t} {\includegraphics[width=0.65\columnwidth]{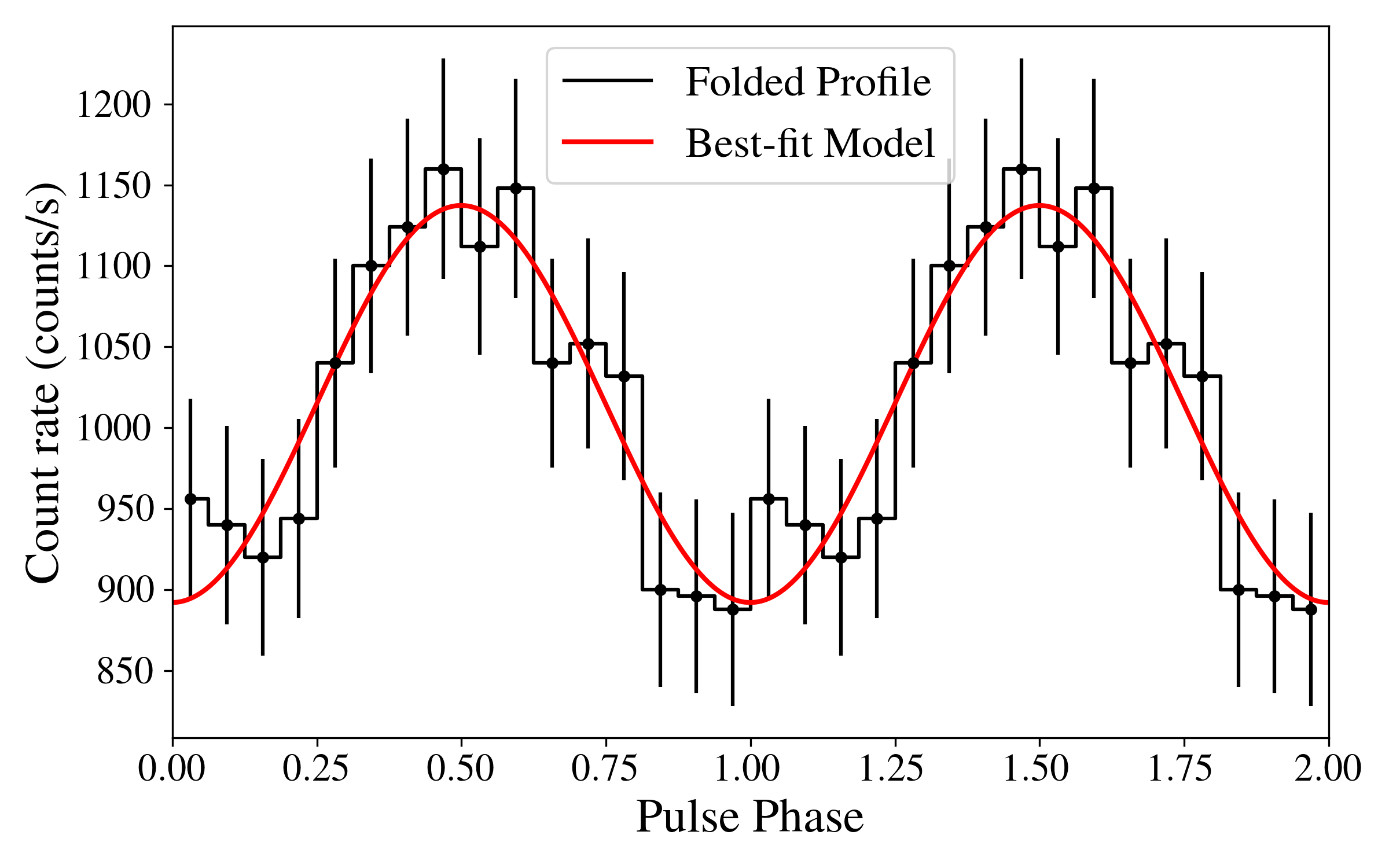}}
  \adjustbox{valign=t} {\includegraphics[width=0.75\columnwidth]{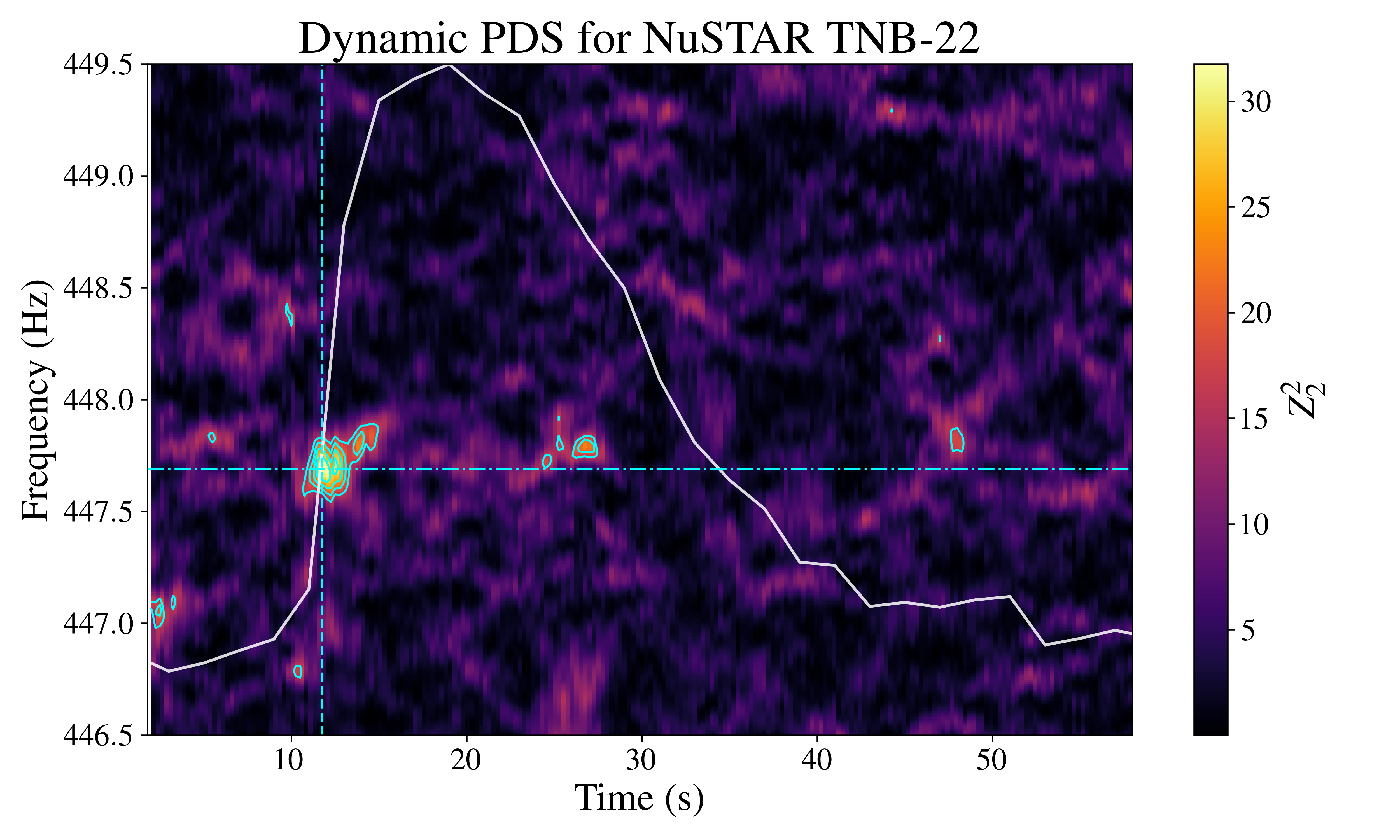}}
   \hspace{0.01\textwidth}
   \adjustbox{valign=t}{\includegraphics[width=0.65\columnwidth]{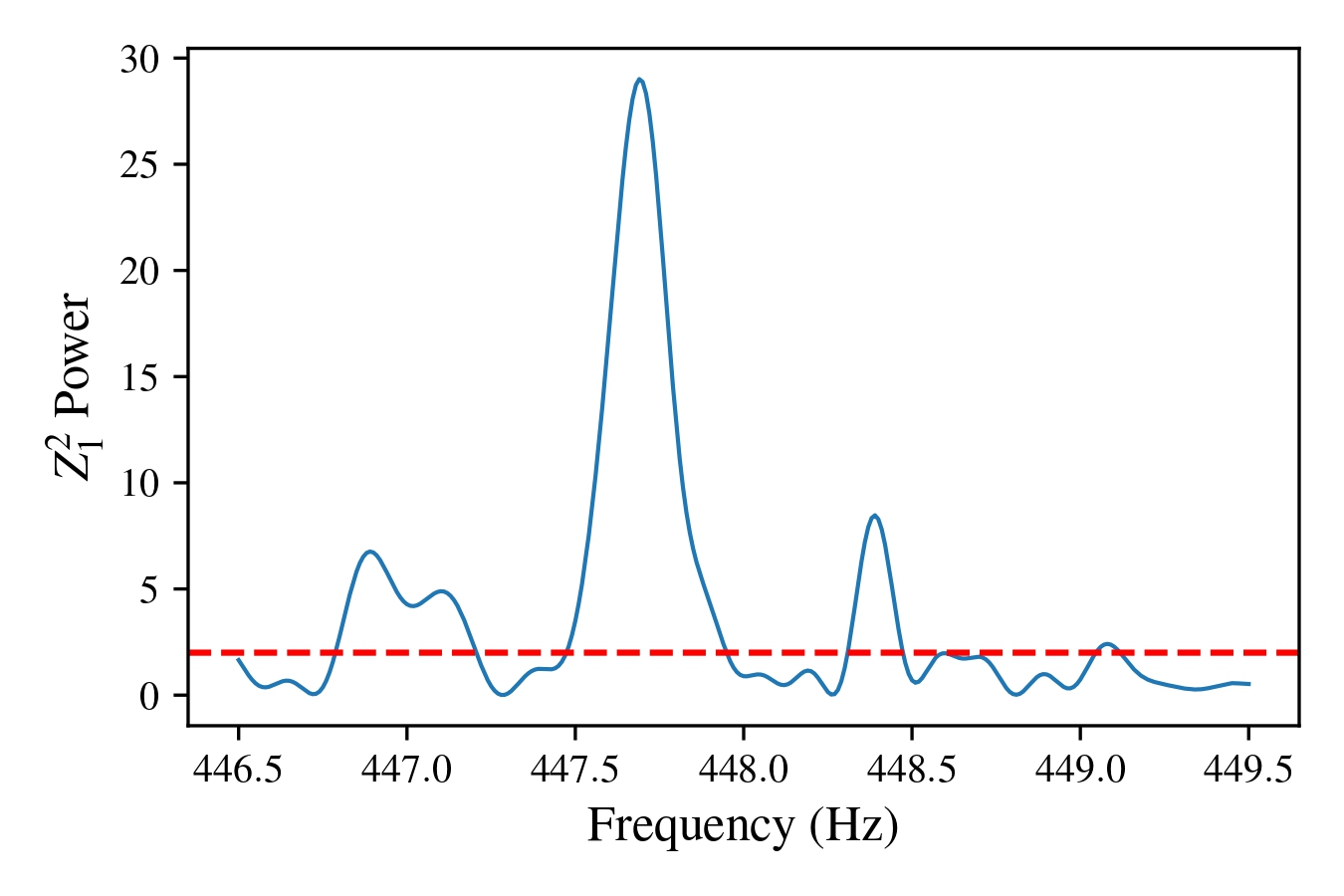}}
    \hspace{0.02\textwidth}
\adjustbox{valign=t} {\includegraphics[width=0.63\columnwidth]{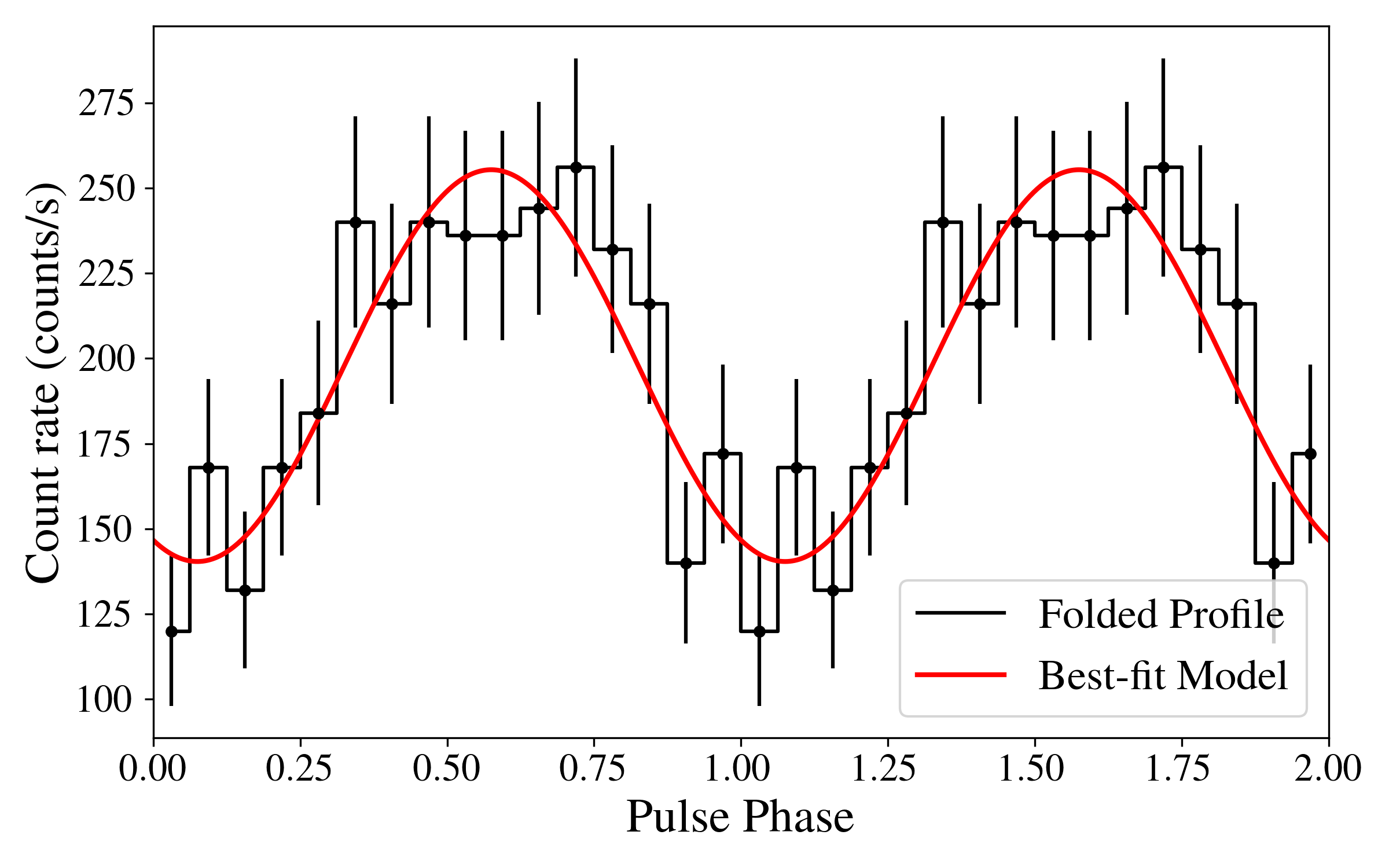}}
\caption{{\bf Left panels:} Dynamic power density spectra of \src{}, overlaid with the \xmm\ EPIC-PN and \nustar\ burst light curves in the top and bottom panels, respectively. Dynamic PDS computed with a 4-s sliding $Z^2$ window (0.25-s step) over a 3 Hz band around the spin frequency at 448 Hz with a resolution of 0.01 Hz. Contours span $Z^2=14-32$ in steps of 4. Cyan dashed lines mark the spin frequency (horizontal) and the time of maximum power (vertical). {\bf Middle panels:} Power density spectra from 8-s and 6-s light-curve segments of the \xmm\ EPIC-PN and \nustar\ bursts, respectively, centered on the maximum power. {\bf Right panels:} The pulse profiles of \src{}, generated by using the burst oscillation frequencies of 448 Hz for \xmm~ and 447.7 Hz for \nustar~ data, are shown along with the best-fit sinusoidal functions.}
    \label{fig:DPDS_XMM}
\end{figure*}
%%%%%%%%%%%%%%%%%%%%%%%%%%%%%%%%%%%%%%%%%%%%%%%%%%%
\section{Results}
\label{res}
In this work, we performed a detailed timing analysis of all thermonuclear bursts detected from the AMXP \src{} in the \xmm~, \nicer{}, and \nustar{} observations during the 2024 X-ray outburst. A total of 39 thermonuclear X-ray bursts were detected in the \nicer{} (4 bursts), \xmm{} EPIC-PN (13 bursts), and \nustar{} (22 bursts) datasets. The extensive, high-quality coverage of these datasets makes them well-suited for investigating burst oscillations, which had not previously been detected from this source. We report the discovery of burst oscillations from \src{} using \xmm{} and \nustar{} observations.
%%%%%%%%%%%%%%%%%%%%%%%%%%%%%%%%%%%%%%%%%%%%%%%%%%%%%
\subsection{Timing analysis: Burst oscillation}
\label{timing}
We conducted a comprehensive timing analysis of all X-ray bursts detected from \src{} during the 2024 observation campaign using \nicer, \xmm, and \nustar, in search of burst oscillations. For \xmm{} and \nicer{}, we used events in the 0.5-10 keV range, while for \nustar{}, we selected the 3-40 keV band. A sliding-window search was performed on each burst using 4~s FFT segments. The search covered a 60-s interval starting approximately 10 s before the onset of the burst and extending to $\sim$50~s after the onset. We computed the $Z^{2}$ statistic over a 3~Hz frequency band centered on the measured spin frequency ($\sim$448 Hz), using sliding windows with a step size of 0.25~s. 

To search for burst oscillations, we computed the $Z_n^2$ statistic \citep{Buccheri1983} directly from the photon arrival times. For a trial frequency $\nu$ and $N$ photons, the $Z_n^2$ statistic is expressed as
\[
Z_n^2 = \frac{2}{N} \sum_{j=1}^{n}
\left[
\left( \sum_{k=1}^{N} \cos(2\pi j \nu t_k) \right)^2
+
\left( \sum_{k=1}^{N} \sin(2\pi j \nu t_k) \right)^2
\right].
\]

We used the first two harmonics, $Z_1^2$ and $Z_2^2$, for the search for burst oscillations. Under the null hypothesis of no signal, $Z_n^2$ follows a $\chi^2$ distribution with $2n$ degrees of freedom, giving the single-trial false–alarm probability (for the first harmonic, n=1)
\[
p = \exp\!\left(-\frac{Z_1^2}{2}\right).
\]

To account for the number of searched time–frequency bins, we computed the multi–trial probability
\[
p_{\rm multi} = 1 - (1 - p)^{N_{\rm trials}},
\]
where $N_{\rm trials}$ is either the full number of evaluated $(t,f)$ bins or the number of statistically independent trials set by the duration of the window. 
 
While searching for oscillations during the thermonuclear bursts observed with \nicer, it is found that significant burst oscillations are not detected in any of the bursts. Considering this, we proceeded with carrying out our investigation using \xmm~ and \nustar~ observations. The light curves from the \xmm~ and \nustar~ observations with 1~s bin time are shown in Figure~\ref{fig:burst_lc}, showing 22 X-ray bursts observed with \nustar~ and 13 bursts with \xmm. Each thermonuclear burst (TNB) is assigned an identification number, as marked in the figure.  

Using \xmm{} EPIC-PN data, we detected significant oscillations (trial-corrected significance$\ge3\sigma$) based on the $Z_2^{2}$ statistic in 7 of the 13 bursts. For six bursts, the oscillations appear significant in both the $Z_{1}^{2}$ and $Z_{2}^{2}$ power spectra, giving high confidence in detection. At the known spin frequency of $\sim447.9$ Hz \citep{Ng24}, we also found a strong candidate burst-oscillation signal in the \xmm\ TNB-3 interval. Using a 4.0 s sliding window stepped by 0.25 s (113 windows), we searched a 3.0 Hz band centered at 448 Hz. The strongest peak reaches $Z_{1}^{2}=30.75$ at 447.99 Hz, corresponding to a single-trial probability of $p_{\rm single}=2.1\times10^{-7}$ ($\approx5.1\sigma$). The full search includes 67,725 raw trials, yielding a raw multi-trial probability of $p_{\rm raw}=1.4\times10^{-2}$ ($\approx2.2\sigma$), which represents a conservative limit given the strong overlap and dependency between segments. To estimate a more realistic significance of the burst oscillation, we consider 12 independent frequency bins per window, corresponding to 180 statistically independent trials. The corresponding independent-trial probability becomes $p_{\rm ind}=3.7\times10^{-5}$ ($\approx3.9\sigma$). A consistent peak is also present in the $Z_{2}^{2}$ spectrum ($Z_{2}^{2}=30.92$ at 447.98 Hz), with $p_{\rm single}=1.9\times10^{-7}$ ($\approx5.1\sigma$), $p_{\rm raw}=1.3\times10^{-2}$ ($\approx2.2\sigma$), and an independent-trial probability of $p_{\rm ind}=3.5\times10^{-5}$ ($\approx4.0\sigma$).

Although the segments overlap in time (step = 0.25 s), the trial correction is based on independent frequency bins, providing a reliable estimate of the significance of the burst oscillation. The burst oscillation characteristics are summarized in Table~\ref{tab:burst_oscillation}. The fractional rms amplitude $A_{\rm rms}$ is defined as
\begin{equation}
A_{\rm rms} = \sqrt{\frac{P_s}{N_m} \left( 1 - \frac{N_{\rm bkg}}{N_m} \right)},
\end{equation}
where $P_s$ is the signal power, $N_m$ is the total number of X-ray photons in a segment, and $N_{\rm bkg}$ is the background count. For bursts with high count rates, the background contribution is negligible ($N_{\rm bkg} \ll N_m$), allowing the approximation
\begin{equation}
A_{\rm rms} \simeq \sqrt{\frac{P_s}{N_m}}
\label{eqn2}.
\end{equation}

The signal power was estimated from the measured Fourier power $P_m$ using the noise–subtraction procedure of \citet{Groth1975}. During the \xmm{} TNB-3 burst, we detected a maximum $Z_2^2$ power of $P_m \simeq 31$ at 448 Hz. The corresponding time window (4-s duration) contains $N_m = 4089$ photons, producing a noise–corrected signal power of $P_s = 26.5 \pm 10.5$. Using Equation~\ref{eqn2}, we obtained a fractional rms amplitude of $A_{\rm rms} = (8.1 \pm 1.6)\%$.

The dynamic power spectrum of \xmm{} EPIC-PN TNB-3 is overlaid with the burst light curve and shown in the top-left panel of Figure~\ref{fig:DPDS_XMM}. The power spectrum with $Z_1^2$ statistics is generated from a 4-s segment (the same segment used in the earlier calculation) around the highest $Z_1^2$ power, and the PDS is shown in the top-middle panel of Figure~\ref{fig:DPDS_XMM}. The pulse profile is now generated by folding the light curve for the same segment using the burst oscillation frequency of $\sim$448 Hz in 16 phase bins. The pulse profile was modeled using a sinusoidal function of the form $F(t) = A + B \, \sin(2 \pi \nu t - \phi_0)$, where A is the constant offset, B is the amplitude, $\nu$ is the oscillation frequency, and $\phi_0$ is the phase. The best-fit model provides A = $1015 \pm 16$, B= $123 \pm 23$, and $\phi_0$= $1.6 \pm 0.18$ radians ($0.50\pm0.06)\pi$. The fractional rms amplitude of the folded pulse profile is estimated to be $f_\textrm{rms}$= $B/\sqrt{2}A$= $8.5 \pm 1.6$ \%, which is consistent with the previous measurement with the $z^2$ power. The best-fit folded pulse profile is shown in the top-right panel of Figure~\ref{fig:DPDS_XMM}.

The \nustar{} data also exhibit significant burst oscillations in 6 of the 22 observed bursts based on the $Z_2^2$ statistic (trial-corrected significance$\ge3\sigma$). In TNB-4, TNB-9 and TNB-22, the signals are significant in both the $Z_{1}^{2}$ and $Z_{2}^{2}$ power spectra, providing robust confidence in these detections. The dynamic power spectrum of \nustar~ TNB-22 is shown in the bottom-left panel of Figure~\ref{fig:DPDS_XMM}. The $Z_1^2$ power spectra and best-fit pulse profile are shown in the bottom-middle and bottom-right panels of Figure~\ref{fig:DPDS_XMM}, respectively. Burst oscillations are detected at $\sim 447.7$ Hz in the \nustar~ data, consistent with the signal seen in the dynamic power spectrum. At this frequency, we detected a peak $Z_1^2 = 29.2$ in the combined FPMA+FPMB data using a 4.0 s sliding window stepped by 0.25 s. The  probability of single-trial is $p_{\rm single} = 4.5 \times 10^{-7}$ ( $\approx 4.9\sigma$). Using 12 independent frequency bins per window and a total of 180 independent trials, the probability of independent trials becomes $p_{\rm ind} = 8.1 \times 10^{-5}$ ($\approx 3.8\sigma$). The raw multi-trial probability across all 67,725 trials is $p_{\rm raw} = 3.0 \times 10^{-2}$ ($\approx1.9\sigma$), which is a conservative estimate, as the segments are correlated, and the true significance of the signal is higher. The $Z_2^2$ spectrum shows a consistent peak of $Z_2^2 = 31.8$ at the same frequency. This corresponds to a single-trial probability of $p_{\rm single} = 1.3 \times 10^{-7}$ ($\approx 5.2\sigma$), an independent-trial probability of $p_{\rm ind} = 2.3 \times 10^{-5}$ ($\approx 4.1\sigma$), and a raw multi-trial probability of $p_{\rm raw} = 8.6 \times 10^{-3}$ ($\approx 2.4\sigma$). The trial correction is based on independent frequency bins, providing a reliable estimate of the significance of burst oscillation. 

The properties of burst oscillations detected with \nustar~ and \xmm~ are summarized in Table~\ref{tab:burst_oscillation}, with representative dynamic power density spectra (DPDS) shown in Figures~\ref{fig:DPDS_NUSTAR_all} and \ref{fig:DPDS_XMM_all}. For each burst, the DPDS were generated to identify intervals with significant oscillation power, and pulse profiles were generated by folding the light curves for segments of excess power at the detected oscillation frequency. The resulting profiles are approximately sinusoidal and were modeled with a sinusoidal function. The fractional rms amplitudes were estimated from both the best-fit model parameters and the $Z^2$ power, which yielded consistent results in all cases.
 
For the \nustar~ burst TNB-22, the maximum $Z^2$ power of $P_m \sim 31$ was detected at 447.7 Hz, corresponding to a noise-corrected signal power of $P_s = 26.5 \pm 10.5$ (Equation~\ref{eqn2}). This yields a fractional rms amplitude of $18.2 \pm 3.6$\%. Sinusoidal modeling of the folded pulse profile gives best-fit parameters $A = 198 \pm 7$, $B = 57 \pm 10$, and $\phi_0 = -1.1 \pm 0.2$ radians ($-0.35 \pm 0.06)\pi$, implying a fractional rms amplitude of $f_{\rm rms} = B/\sqrt{2}A = 20.5 \pm 3.6$\%, consistent with the $Z^2$-based estimate. A similar analysis was performed for all bursts observed with \xmm~ EPIC-PN. The detected oscillation frequencies, significances, and fractional rms amplitudes are listed in Table~\ref{tab:burst_oscillation}, with rms values in the range 7–12\%. The corresponding dynamic power spectrum and best-fit pulse profiles are shown in Figure~\ref{fig:DPDS_XMM_all}.

In addition, for the X-ray bursts observed simultaneously with \xmm\ and \nustar, we generated the dynamic power spectra using \nustar~ in the 3--8 keV energy range. Among the nine common bursts, \xmm~ data show significant burst oscillations (trial-corrected significance in $Z_2^2$ test $\geq 3\sigma$) in three of them. Of these, oscillations are detected in two corresponding \nustar\ bursts: \xmm\ TNB-3 and TNB-12, matched with \nustar\ TNB-13 and TNB-19, respectively. As the instruments have different effective areas, burst emission may be dominated in soft X-rays below $\sim$6 keV. As a result, some oscillation signals may fall below \nustar~ sensitivity and remain undetected. Therefore, any direct comparison between the two datasets should be treated with caution.
%%%%%%%%%%%%%%%%%%%%%%%%%%%%%%%%%%%%%%%%%%%%%%%%%%%%%%%%%%%%%%5
\subsubsection{Pulsation before the burst onset}
\begin{figure}
     \includegraphics[width=\columnwidth]
 {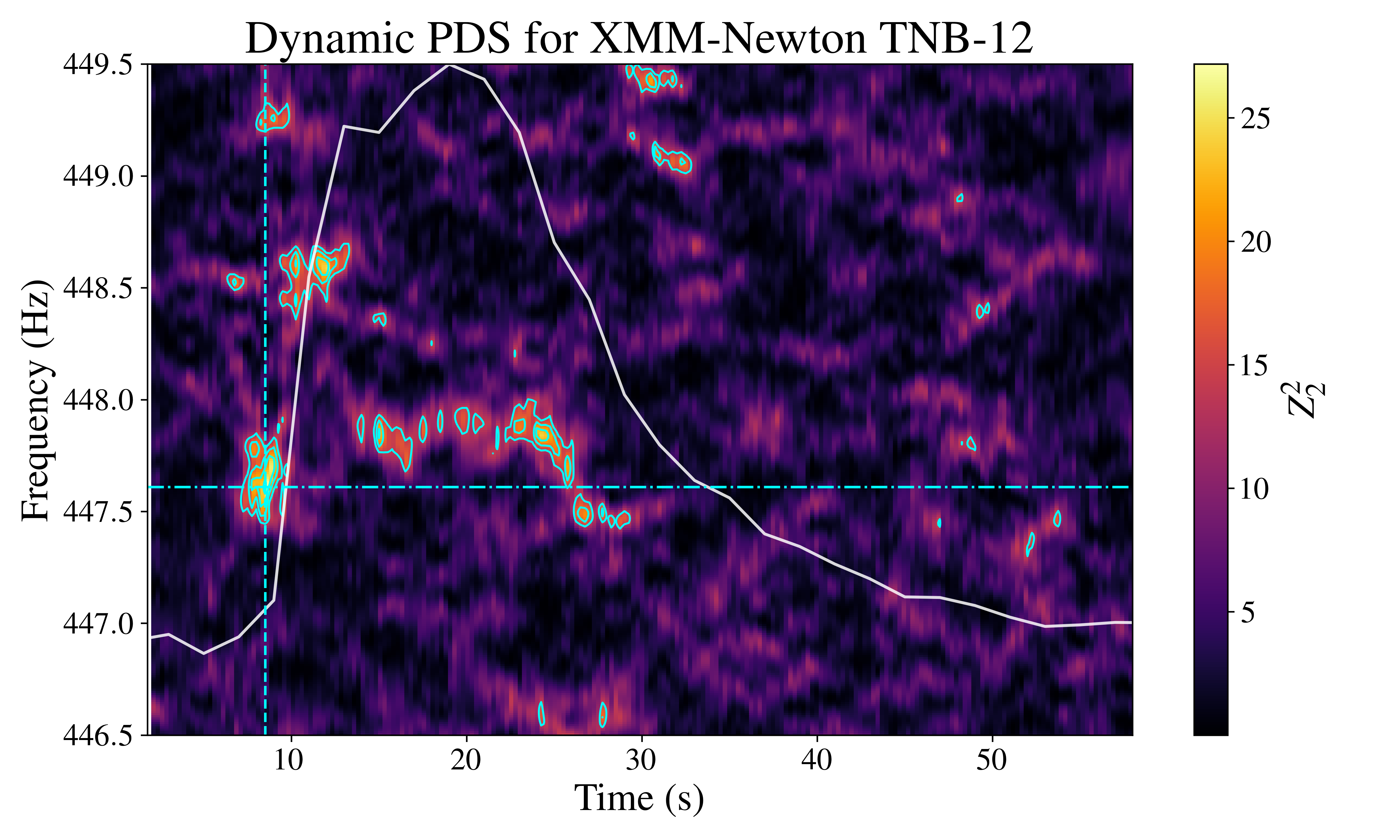}
   \includegraphics[width=0.95\columnwidth] {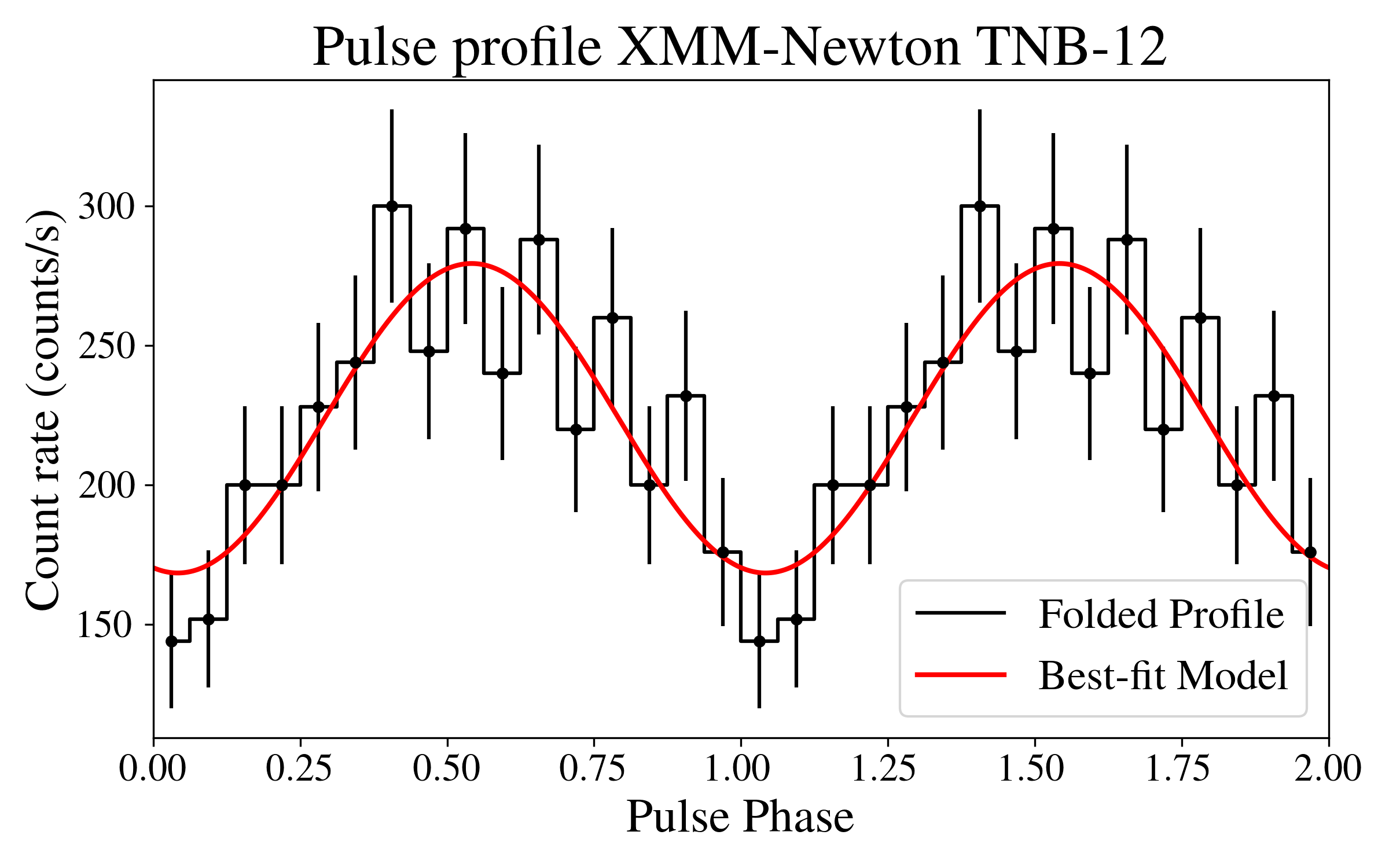}
\caption{The dynamic power spectrum and the best-fit pulse profiles for the segment are shown with maximum power before the burst for \xmm~ TNB-12.}
    \label{fig:DPDS_XMM_prior}
\end{figure}

In our study, we also found that pulsations are present immediately before the burst onset, as observed with \xmm~ EPIC-PN. In \xmm~TNB-12, we detect burst oscillations just before the start of the burst, which is, to our knowledge, the second source for which burst oscillations have been observed just before the burst. Figure~\ref{fig:DPDS_XMM_prior} shows the dynamic power spectrum and the corresponding best-fit pulse profile during the interval immediately preceding the burst onset. A coherent oscillation is detected at 447.61 Hz, with a maximum $Z^2$ power of $\sim27$. The single-trial probability is $1.3\times10^{-6}$ (corresponding to $\simeq4.7\sigma$), while the trial-corrected significance is 3.5$\sigma$. The pulse profile is modeled using a sinusoidal function, and the rms amplitude is estimated to be $17.5\pm3.3$\%. The oscillations reappear during the peak and decay phases.
%%%%%%%%%%%%%%%%%%%%%%%%%%%%%%%%%%%%%%%%%%%%%%%%%%%%%
%%%%%%%%%%%%%%%%%%%%%%%%%%%%%%%%%%%%%%%%%%%
%%%%%%%%%%%%%%%%%%%%%%%%%%%%%%%%%%%%%%%%%%%%%%%%%%%%%
\begin{figure*}
     \includegraphics[width=\columnwidth] {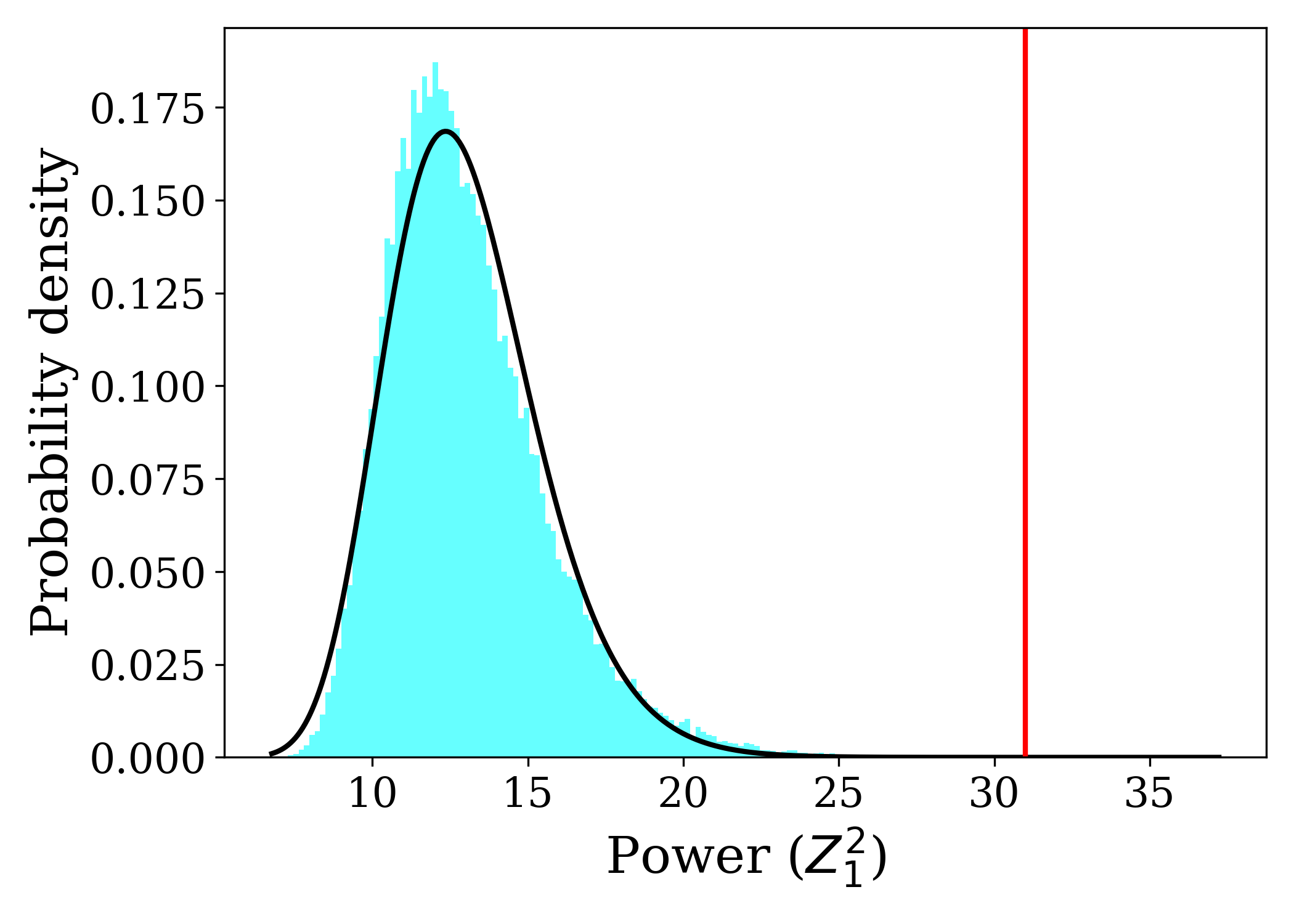}
      \includegraphics[width=\columnwidth] {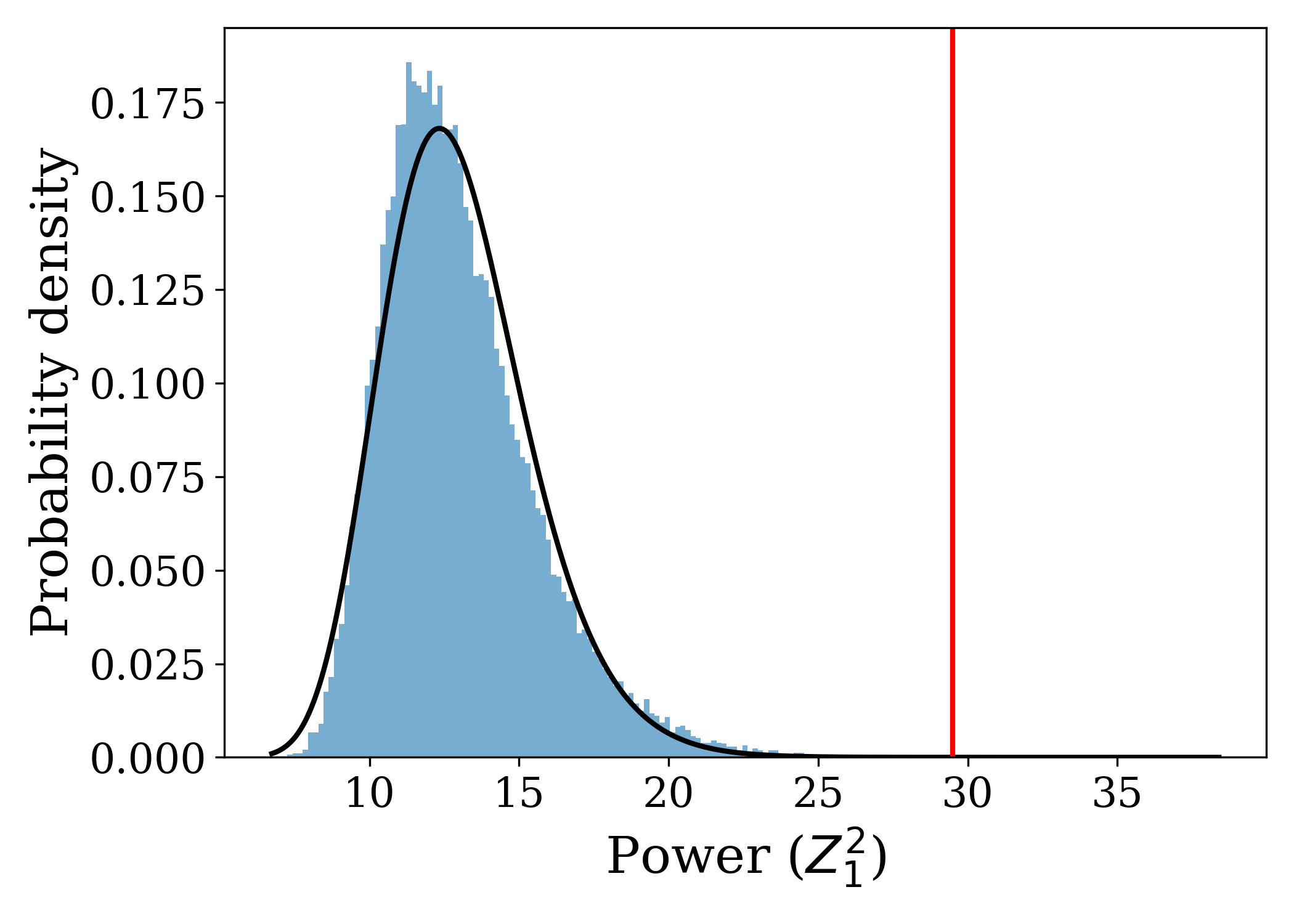}
\caption{Distribution of maximum $Z_1^2$ values from 50,000 Monte Carlo simulations of the burst. The black curve shows the posterior log-normal fit, while the red vertical line marks the observed maximum $Z_1^2$ from the \xmm~ EPIC-PN TNB-3 burst (left) and \nustar~ TNB-22 (right).}
    \label{fig:MCMC_XMM}
\end{figure*}
%%%%%%%%%%%%%%%%%%%%%%%%%%%%%%%%%%%%%%%%%%%
%%%%%%%%%%%%%%%%%%%%%%%%%%%%%%%%%%%%%%%%%%%%%%%%%%%%%%%%%%%%%%
\subsubsection{Simulations of X-ray burst light curves}
The significance of the candidate oscillations observed during the X-ray bursts is determined using the Monte Carlo simulation of light curves. The significance of the burst oscillation was assessed using overlapping timing windows, which are not statistically independent. To account for the nonstationary nature of the burst light curve and the biases introduced by overlapping windows, we estimated the false-alarm probability through numerical simulations. We applied the same timing-analysis procedure to the simulated burst profiles as to the real data. For each burst, we determined how often the simulations produced a maximum $Z^2$ value equal to or greater than the observed value. From the resulting distribution of simulated maximum $Z^2$ values, we computed the p-value for each burst and converted these p-values into their corresponding $\sigma$ significances. 

Following the methods described in \citet{Bilous2019}, \citet{Li2022}, and \citet{Bult2021}, we generated a large ensemble of synthetic burst light curves by randomizing photon arrival times according to the observed burst count-rate profiles. This approach preserves the overall burst morphology and Poisson counting statistics, thereby providing an appropriate realization of the null hypothesis. Each simulated light curve was analyzed using the same burst-oscillation search procedure described in Section~\ref{timing}. We performed 50,000 Monte Carlo simulations to estimate the distribution of maximum $Z_1^2$ powers under the null hypothesis. From $5 \times 10^{4}$ simulations, none yielded a maximum power exceeding the observed value. The observed maximum power, $P_m = 31$, is highly significant. The single trial probability, the posterior probability from the lognormal MCMC fit, and the empirical Monte Carlo p-value are estimated to be $1.8\times10^{-7} (\simeq 5.1\sigma)$, $p_{\rm post} = 1.3 \times 10^{-6}$ ($\simeq 4.7\sigma$), and $p_{\rm emp} = 8.0 \times 10^{-5}$ ($\simeq3.8\sigma$), respectively. The left panel of Figure~\ref{fig:MCMC_XMM} shows the log-normal fit to the simulated maxima, with the observed power marked for \xmm~ TNB-3, confirming the burst oscillation is highly unlikely to arise by chance. 

Similarly, we carried out 50,000 Monte Carlo simulations to estimate the null distribution of the maximum $Z_1^2$ power for the \nustar\ data and modeled the resulting maxima using a log-normal MCMC framework. The observed maximum power was 29.5 in TNB-22, and none of the simulated results exceeded the maximum observed power, which indicates the observed power is not a random fluctuation. The observed maximum power is highly significant under all statistical assessments. The empirical Monte Carlo p-value is $p_{\rm emp} = 2.2 \times 10^{-4}$ ($\simeq3.5\sigma$) for the \nustar~ TNB-22. The posterior distribution obtained from the log-normal MCMC fit yields a much smaller median probability, $p_{\rm post} = 4.9 \times 10^{-6}$ ($\simeq4.4\sigma$). The right panel of Figure~\ref{fig:MCMC_XMM} shows the log-normal fit to the simulated maxima, with the observed \nustar\ TNB-22 power marked, confirming that the detected burst oscillation is unlikely to arise from random noise.
%%%%%%%%%%%%%%%%%%%%%%%%%%%%%%%%%%%%%%%%%
\begin{table*}[t]
\centering
\scriptsize
\caption{Details of burst oscillations from \src{} using \xmm{} and \nustar{} observations. The parameters are estimated from the dynamic PDS, which is generated using a 4-s sliding window with the $Z^2$ statistic and a step size of 0.25 s over a 3 Hz band centered around the spin frequency. A frequency resolution of 0.01 Hz is employed.}
\begin{tabular}{lcccccccc}
\hline
Burst no. & $\nu$ (Hz) & $Z^2$ power &$p_\textrm{single}$ & Sig. &$p_\textrm{ind}$ & Sig. &\multicolumn{2}{c}{RMS (\%)} \\ 
\cline{8-9}
& & & & & & &$Z^2$ & Profile model \\
\hline
 \multicolumn{9}{c}{\xmm~ EPIC-PN} \\
\hline
TNB-2$^{\star}$ &   447.80 & 31.2 & $1.7\times10^{-7}$ & 5.1$\sigma$ & $3.0\times10^{-5}$  & 4.0$\sigma$ & $8.3\pm1.6$ & $8.6\pm1.6$\\
TNB-3$^{\star}$ & 447.98 & 31.0 & $1.9\times10^{-7}$ & 5.1$\sigma$ & $3.5\times10^{-5}$  & 4.0$\sigma$ & $8.1\pm1.6$ & $8.8\pm1.6$\\
TNB-4$^{\star}$ &   447.78 & 29.8 & $3.3\times10^{-7}$ & 5.0$\sigma$ & $5.5\times10^{-5}$  & 3.9$\sigma$ &$7.7\pm1.5$ & $7.2\pm1.5$ \\
TNB-5$^{\star}$ &   448.00 & 33.0 & $6.7\times10^{-8}$ & 5.3$\sigma$ & $1.2\times10^{-5}$  & 4.2$\sigma$ & $8.3\pm1.6$ & $8.2\pm1.6$\\
TNB-9$^{\star}$ &   447.93 & 28.1 & $7.8\times10^{-7}$ & 4.8$\sigma$ & $1.4\times10^{-4}$  & 3.6$\sigma$ & $7.5\pm1.6$ & $7.8\pm1.5$ \\
Before TNB-12$^{\star}$ &   447.61 & 27.2 & $1.3\times10^{-6}$ & 4.7$\sigma$ & $2.3\times10^{-4}$  & 3.5$\sigma$ & $15.9\pm3.4$ & $17.5\pm3.3$ \\
TNB-12$^{\dagger}$ &   447.85 & 24.1 & $5.8\times10^{-6}$ & 4.4$\sigma$ & $1.0\times10^{-3}$  & 3.1$\sigma$ & $6.2\pm1.2$ & $5.1\pm1.0$ \\
TNB-13$^{\star}$ &   447.93 & 27.6 & $9.9\times10^{-7}$ & 4.8$\sigma$ & $1.8\times10^{-4}$  & 3.6$\sigma$ & $9.9\pm2.1$ & $10.7\pm3.2$\\
\hline
 \multicolumn{9}{c}{\nustar~} \\
\hline
TNB-4$^{\star}$ & 447.85 &  30.7 & $2.1\times10^{-7}$ & 5.1$\sigma$ & $3.8\times10^{-5}$ & 3.9$\sigma$ &  $12.9\pm2.5$ & $11.8\pm2.2$ \\
TNB-7$^{\dagger}$ & 447.87 &  25.8 & $2.5\times10^{-6}$ & 4.6$\sigma$ & $4.4\times10^{-4}$ & 3.3$\sigma$ & $12.2\pm2.7$ & $10.5\pm2.4$ \\
TNB-9$^{\star}$ & 447.96 &  24.8 & $4.0\times10^{-6}$ & 4.5$\sigma$ & $7.3\times10^{-4}$ & 3.2$\sigma$ & $11.6\pm2.6$ & $9.3\pm2.1$ \\
TNB-10 & 447.88 &  22.8 & $1.1\times10^{-5}$ & 4.2$\sigma$ & $1.9\times10^{-3}$ & 2.9$\sigma$ & $15.8\pm3.7$ & $23.1\pm5.5$\\
TNB-13$^{\dagger}$ & 447.86 &  23.8 & $6.6\times10^{-6}$ & 4.4$\sigma$ & $1.2\times10^{-3}$ & 3.0$\sigma$ & $14.8\pm3.4$ & $11.3\pm2.5$\\
TNB-18$^{\dagger}$ & 447.87 &  25.7 & $2.6\times10^{-6}$ & 4.6$\sigma$ & $4.7\times10^{-4}$ & 3.3$\sigma$ & $11.7\pm2.6$ & $12.2\pm2.5$\\
TNB-19 & 447.87 &  22.1 & $1.6\times10^{-5}$ & 4.2$\sigma$ & $2.9\times10^{-3}$ & 2.8$\sigma$ & $15.5\pm3.8$ & $13.2\pm3.7$\\
TNB-22$^{\star}$ & 447.69 &  31.8 & $1.3\times10^{-7}$ & 5.2$\sigma$ & $2.3\times10^{-5}$ & 4.1$\sigma$ & $18.2\pm3.6$ & $20.5\pm3.6$ \\
\hline
\label{tab:burst_oscillation}
	\end{tabular}
 \tablecomments{
$^{\dagger}$ indicates a burst-oscillation with significance $\ge3\sigma$ (single-trial and independent-trial) using only the $Z_2^2$ test.  
$^{\star}$ indicates burst–oscillation with significance $\ge3\sigma$ (single-trial and independent-trial) using both the $Z_1^2$ and $Z_2^2$ tests.
}
\end{table*}

%%%%%%%%%%%%%%%%%%%%%%%%%%%%%%%%%%%%%%%%%%%%%%%%%
\section{Discussion and Conclusions}
\label{dis}
Following its discovery in 2024, the AMXP \src{} has been extensively monitored with \xmm{}, \nicer{}, and \nustar{}. During these observations, 39 thermonuclear X-ray bursts were detected, with several captured simultaneously by multiple instruments. This extensive and high-quality burst sample makes \src{} a very useful source for investigating burst oscillations, which had not been detected from this source before. We report the results of a comprehensive timing analysis of every thermonuclear burst from \src~ observed with the three missions. 

For the first time, we detect burst oscillations from \src{} at 447.7--448.0 Hz, which is consistent with the NS spin frequency. These oscillations are independently detected in both the \xmm{} and \nustar{} datasets, demonstrating their robustness across instruments. In the \xmm{} observation, the burst oscillations are detected in a significant fraction of bursts, often with power in both the $Z_{1}^{2}$ and $Z_{2}^{2}$ statistics, indicating coherent signals. The oscillation strengths, pulse-profile morphology, and fractional rms amplitudes are consistent with those observed in other AMXPs and neutron-star LMXBs \citep{Ga08}. Independent estimates of the fractional rms amplitude derived from pulse-profile modeling and from the $Z^{2}$ statistic are in good agreement, further supporting the reliability of detections.

In addition, oscillation is detected at 447.61 Hz just before the burst onset as observed in \xmm~ TNB-12. The significance of the signal with a single trial was 4.7$\sigma$, though the trial-corrected significance was 3.5$\sigma$. Pre-rise oscillations may arise from an evolving hot spot, with a weak, slowly growing burst phase too subtle for the count rate but still producing pulsations. Although this contrasts with the rapid onset expected for hydrogen-poor bursts \citep{Cumming2000}, multidimensional effects such as confinement or finite-time flame spreading could explain it. Earlier, \citet{Bostanci2023} reported the first detection of burst oscillations from the bursting LMXB~4U~1728-34 just before the burst onset, observed in two bursts, where the oscillations disappear at the start of the burst rise. The oscillations just before the burst onset can be explained by a small, slowly growing hot spot from off-equator ignition that produces detectable pulsations before the global burst rise becomes visible in terms of count rate \citep{Cavecchi2019, Spitkovsky2002}. Once the flame rapidly spreads across the surface of the neutron star, the emission becomes more symmetric, causing the oscillations to disappear \citep{Spitkovsky2002,Bostanci2023}. In the case of \src{}, oscillations are detected both prior to burst onset and during the burst (\xmm~ TNB-12), placing it among the second sources after 4U~1728-34 \citep{Bostanci2023} reported to exhibit pre-burst oscillations to date. The \nustar{} observation further confirms the presence of coherent burst oscillations in multiple bursts, with pulse profiles that are well described by a sinusoidal function and fractional rms amplitudes comparable to those seen in other bursting neutron-star systems. The consistency between amplitudes inferred from timing statistics and from pulse-profile modeling strengthens the case that these oscillations arise from genuine surface brightness asymmetries. 

The physical mechanism responsible for the brightness asymmetries that give rise to burst oscillations is not yet fully understood. These asymmetries likely arise from localized regions on the neutron-star surface where thermonuclear burning ignites and spreads, but the detailed geometry and propagation of these hot spots remain uncertain. Nevertheless, burst oscillations provide important information about thermonuclear burning and the composition and structure of the outer layers of the neutron star \citep{Strohmayer2006}. In many neutron star LMXBs (e.g., 4U~1728-34, 4U~1636-536, KS~1731-260, X~1658-298), burst-oscillation frequencies are observed to drift by a few Hertz during the burst evolution \citep{Watts2012, Muno2002, Bostanci2023}, with the largest reported drifts reaching $\sim$5~Hz in X~1658-298 \citep{Wijnands2001}. This frequency drift is commonly interpreted as the result of angular-momentum conservation in the expanding and subsequently contracting burning layer, or from changes in the pattern speed of the surface brightness asymmetry \citep{Watts2012, Strohmayer1997}. In contrast, no significant frequency drift is detected in the burst oscillations from \src{} in either the \xmm{} or \nustar{} observation, suggesting a comparatively stable oscillation pattern throughout the burst evolution. Burst oscillations are observed at different stages of thermonuclear X-ray bursts and are thought to arise from distinct physical mechanisms during burst rise, peak, and decay, including localized ignition and flame spreading \citep{Strohmayer1997}, rotational modulation of asymmetric surface emission \citep{Strohmayer2006, Watts2012}, and persistence of surface brightness asymmetries due to hydrodynamic instabilities or global ocean modes \citep{Spitkovsky2002, Cumming2000}.

The absence of burst oscillations in the \nicer~ observations may be explained in the context of mass accretion rate rather than a change in the underlying mechanism. The \nicer~ data were obtained near the X-ray outburst peak, at higher persistent flux and inferred accretion rate than the \xmm~ and \nustar~ observations taken during the decay phase. Such variations in accretion rate may affect the initial burning conditions and the detectability of burst oscillations. It remains unclear why some neutron stars in AMXPs exhibit burst oscillations, while others do not. Even in sources that do show burst oscillations, they are not present in every burst. In some cases, they appear as rarely as once in fourteen bursts (e.g., 4U~1916-053; \citet{Ga08}). Although the high oscillation amplitudes seen early in some bursts fit well with a spreading hot-spot model (e.g., \citet{Strohmayer1997}), explaining oscillations in the burst tail is more difficult, since by that stage the burning should have enveloped the entire stellar surface. These later-time oscillations may instead arise from brightness asymmetries produced by hydrodynamic instabilities \citep{Spitkovsky2002} or from modes excited in the NS ocean \citep{Cumming2000}.
%%%%%%%%%%%%%%%%%%%%%%%%%%%%%%%%%%%%%%%%%%%%%%%%%%%%%%%%%%%%%%%%%%%%%%%%
%%%%%%%%%%%%%%%%%%%%%%%%%%%%%%%%%%%%%%%%%%%%%%%%%%%%%%%%%%%%%%%%%%%%%%%%
\facilities{ADS, HEASARC, \nicer{}, \xmm{}, \nustar{}}

\software{HEASoft V6.33.2 \citep{heasoft}, XSPEC V12.13.0 \citep{Ar96}}, NumPy and SciPy \citep{virtanen20}, Matplotlib \citep{hunter07}, IPython \citep{perez07}. 
%%%%%%%%%%%%%%%%%%%%%%%%%%%%%%%%%%%%%%%%%%%%%%%%%%%%%5
\section*{Acknowledgements}
We thank the anonymous referee for valuable comments that helped to improve the manuscript. The research work at the Physical Research Laboratory, Ahmedabad, is funded by the Department of Space, Government of India. This research has made use of data obtained with \nustar{}, a project led by Caltech, funded by NASA, and managed by NASA/JPL, and has utilized the {\tt NUSTARDAS} software package, jointly developed by the ASDC (Italy) and Caltech (USA). We acknowledge the use of public data from the \xmm{},  \nicer{}, and \nustar{} data archives. 
%%%%%%%%%%%%%%%%%%%%%%%%%%%%%%%%%%%%%%%%%%%%%%%%%%
\section*{Data Availability}
The data used for this article are publicly available in the High Energy Astrophysics Science Archive Research Centre (HEASARC).%\footnote{https://heasarc.gsfc.nasa.gov/db-perl/W3Browse/w3browse.pl}.
%%%%%%%%%%%%%%%%% APPENDICES %%%%%%%%%%%%%%%%%%%%%
%\clearpage
\appendix

\begin{figure*}
 \includegraphics[width=0.35\columnwidth] {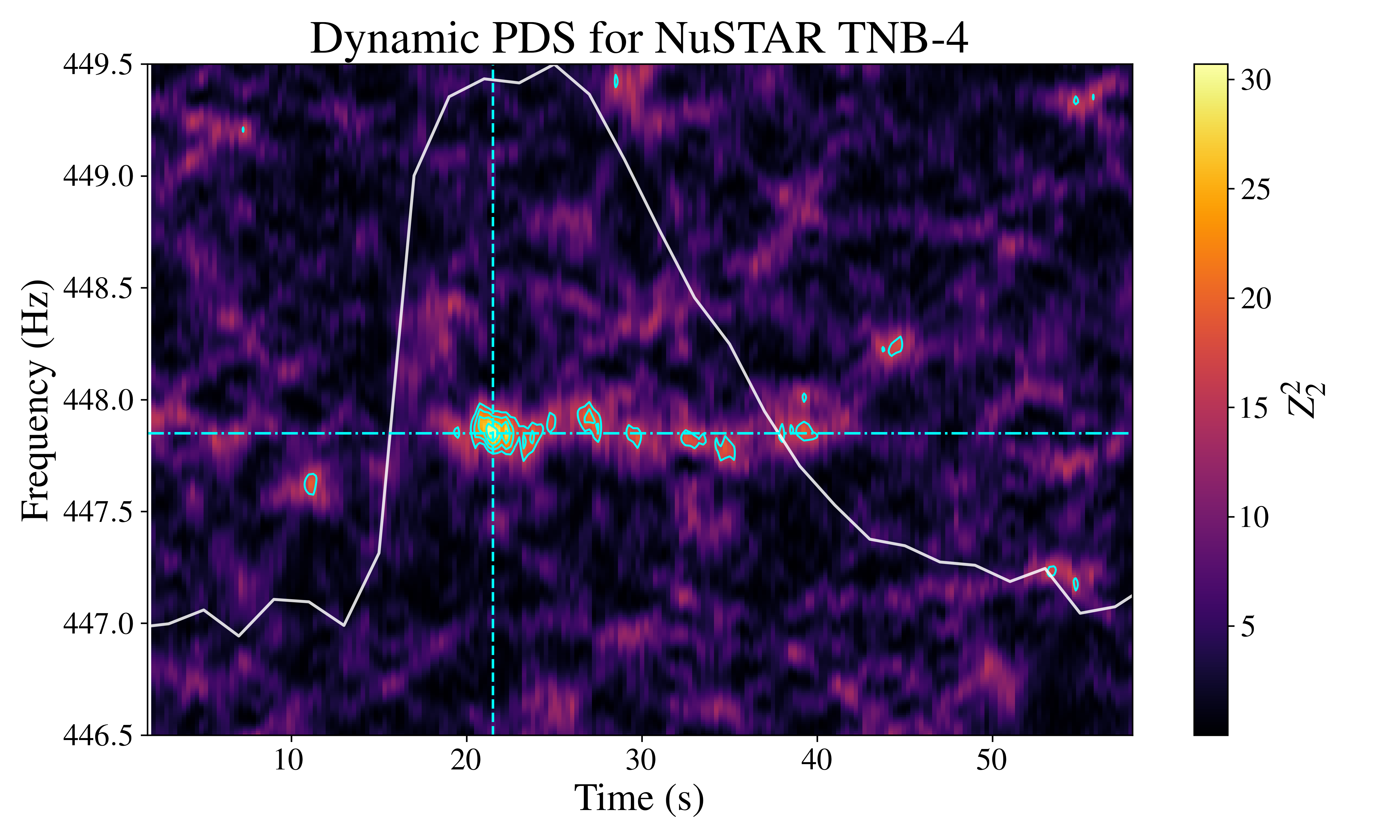}
    \hspace{-0.02\textwidth}
  \includegraphics[width=0.35\columnwidth]{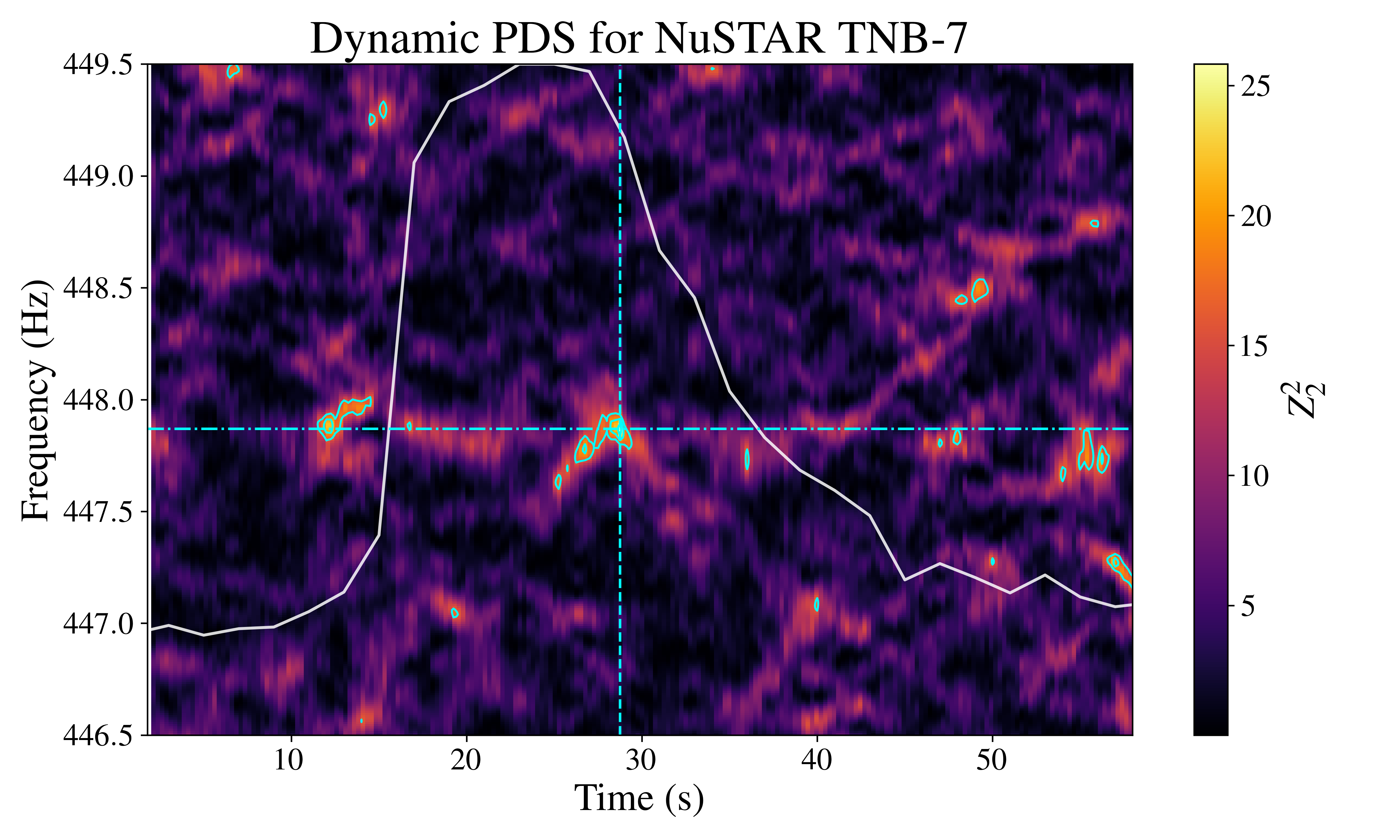}
    \hspace{-0.02\textwidth}
 \includegraphics[width=0.35\columnwidth] {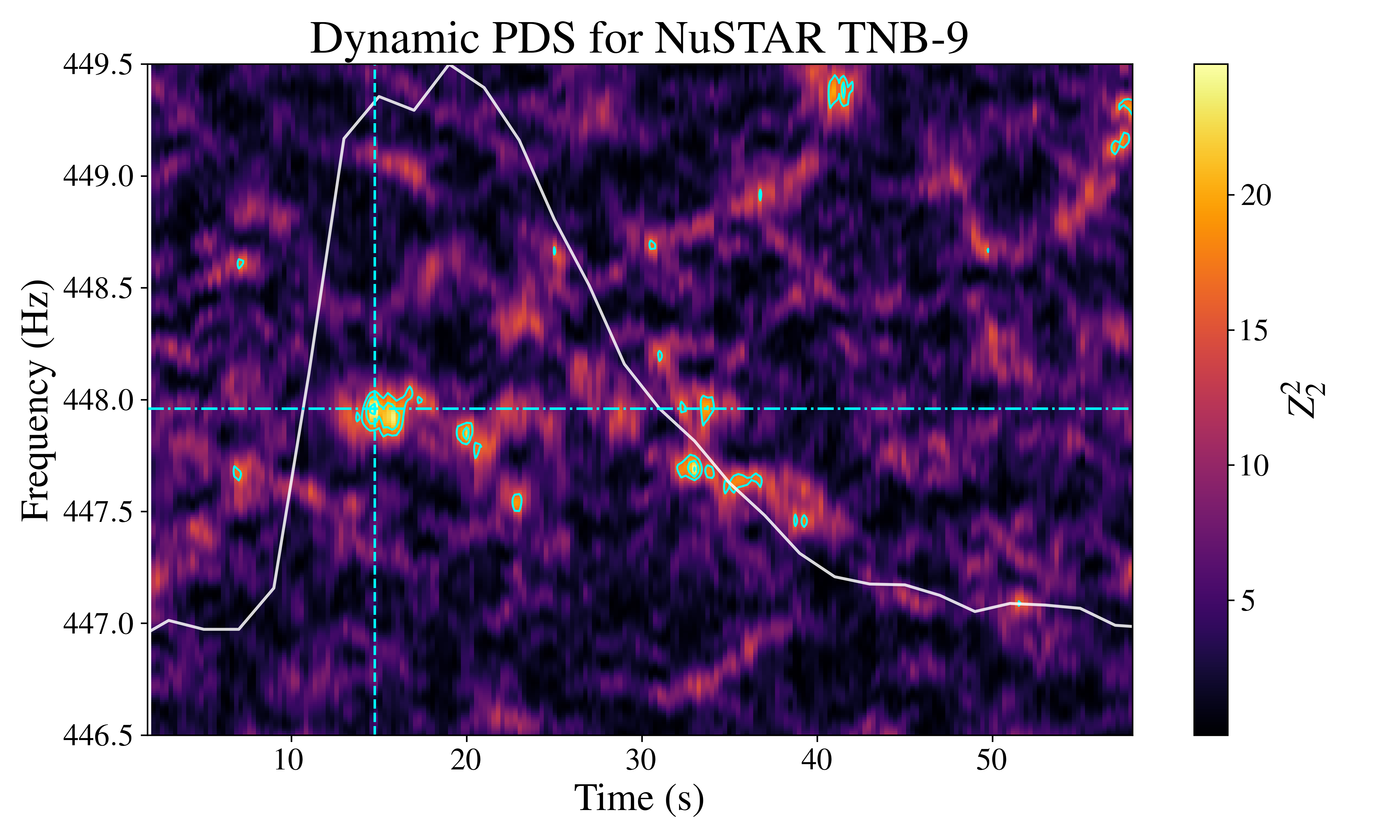}
 { \includegraphics[width=0.3\columnwidth] {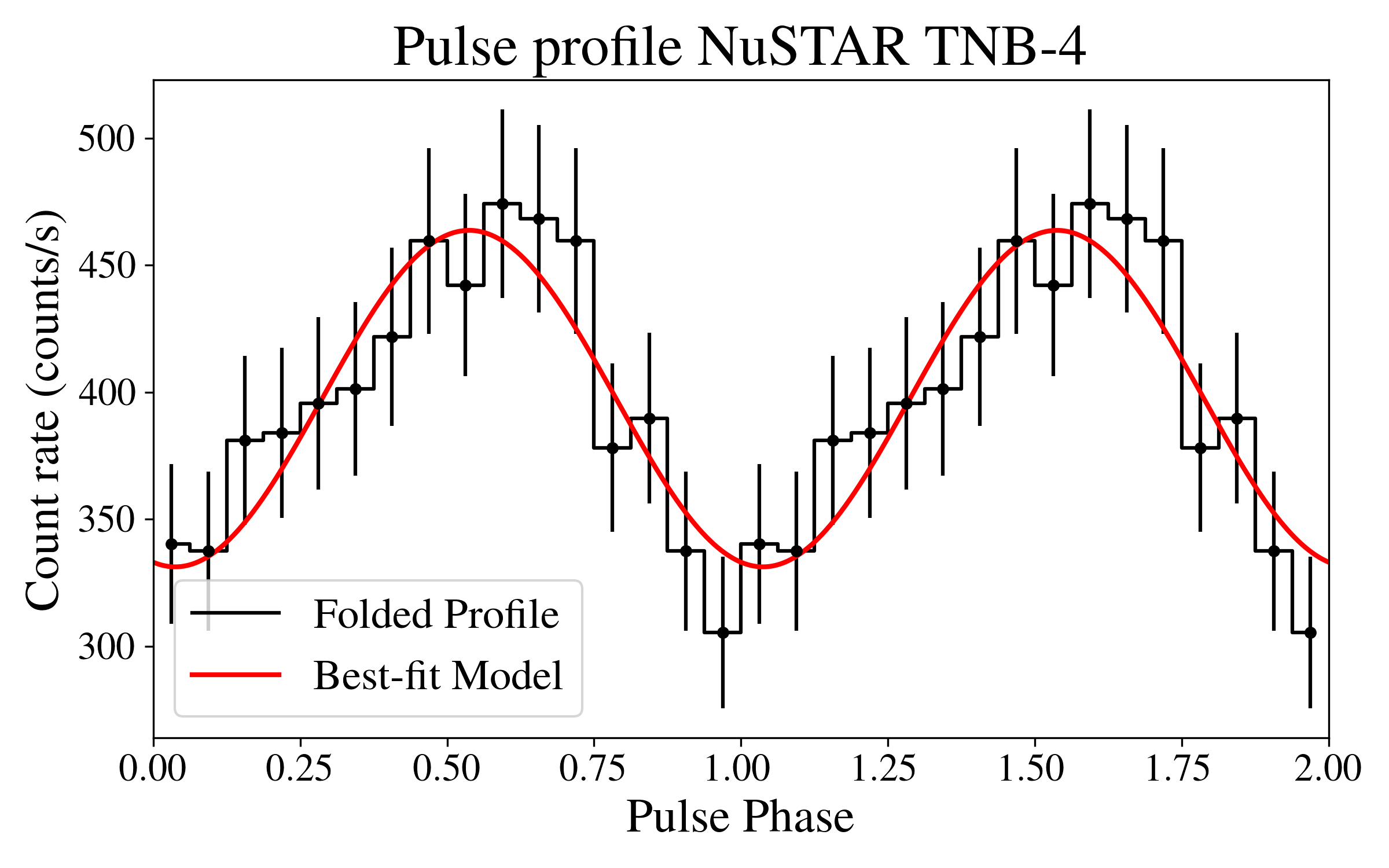}}
  { \includegraphics[width=0.3\columnwidth] {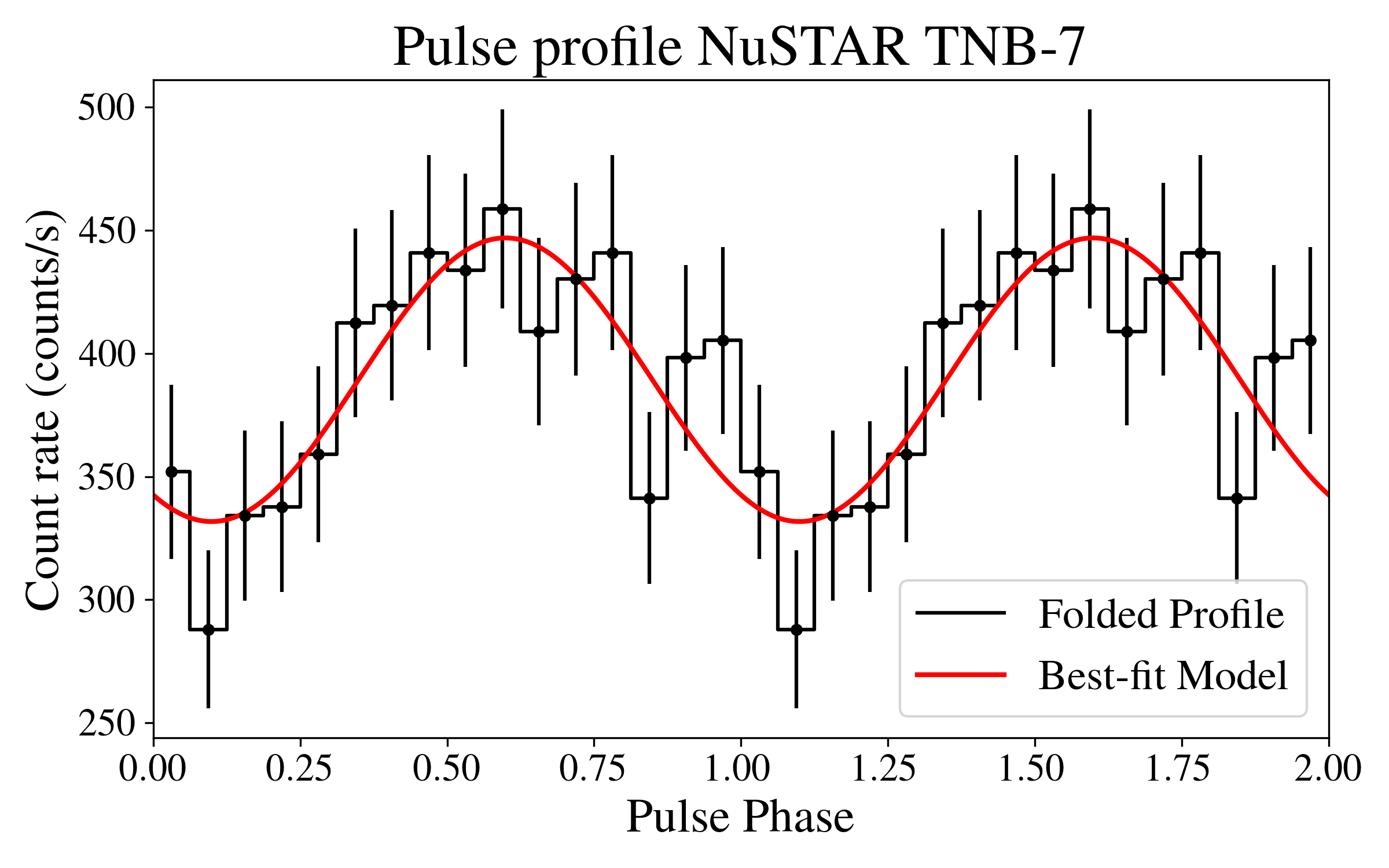}}
  { \includegraphics[width=0.3\columnwidth] {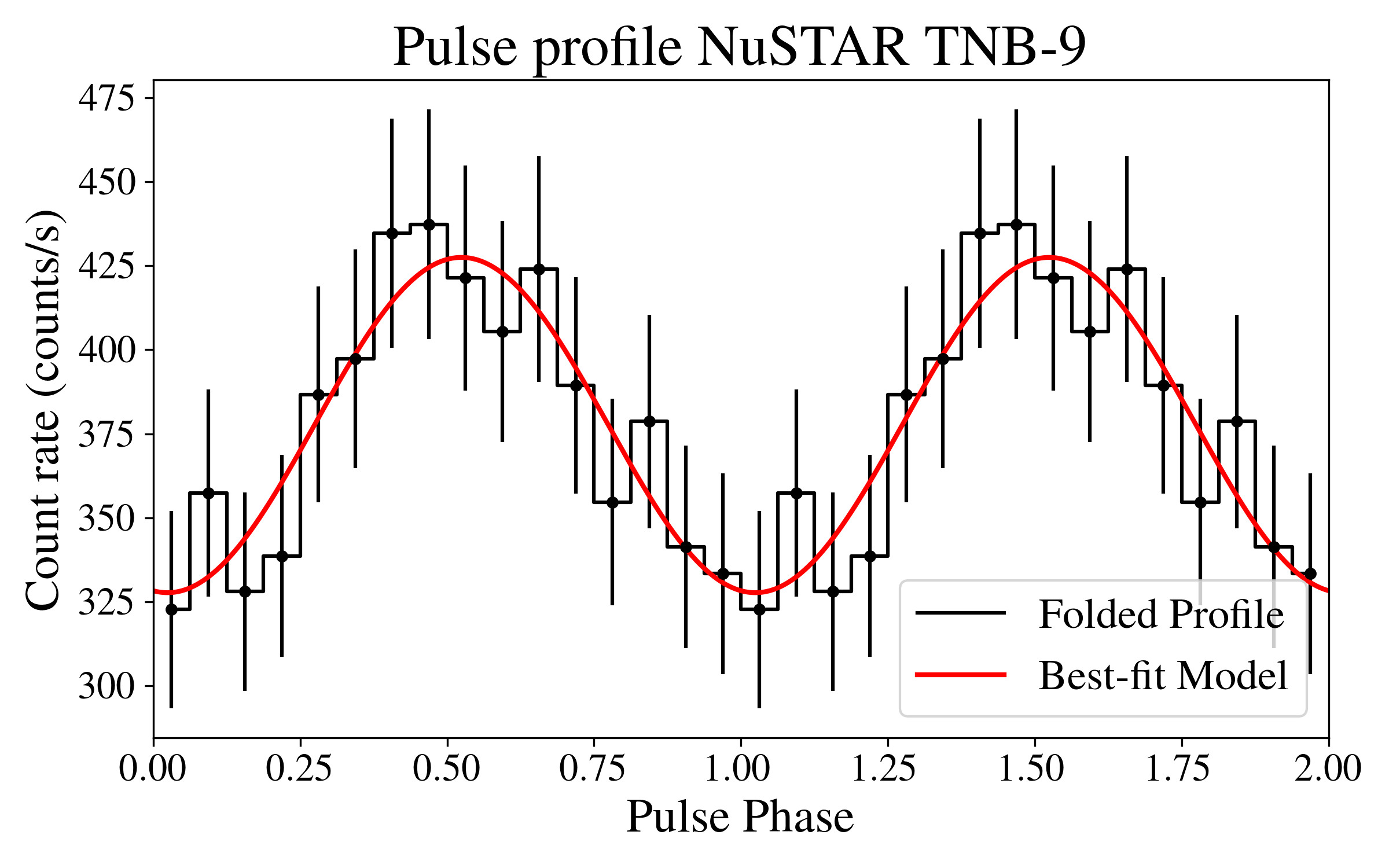}}
     \includegraphics[width=0.35\columnwidth] {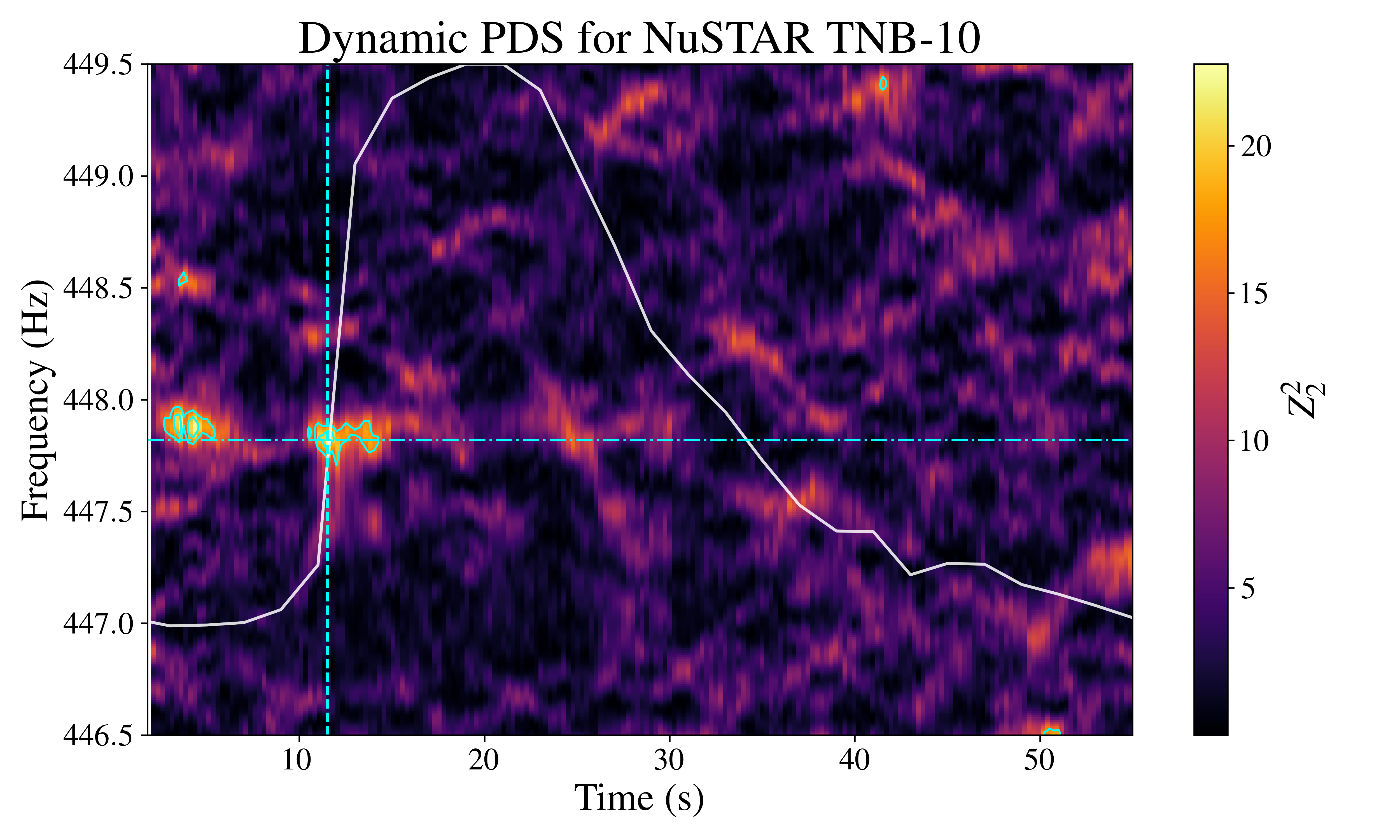}
      \hspace{-0.02\textwidth}
   \includegraphics[width=0.35\columnwidth]{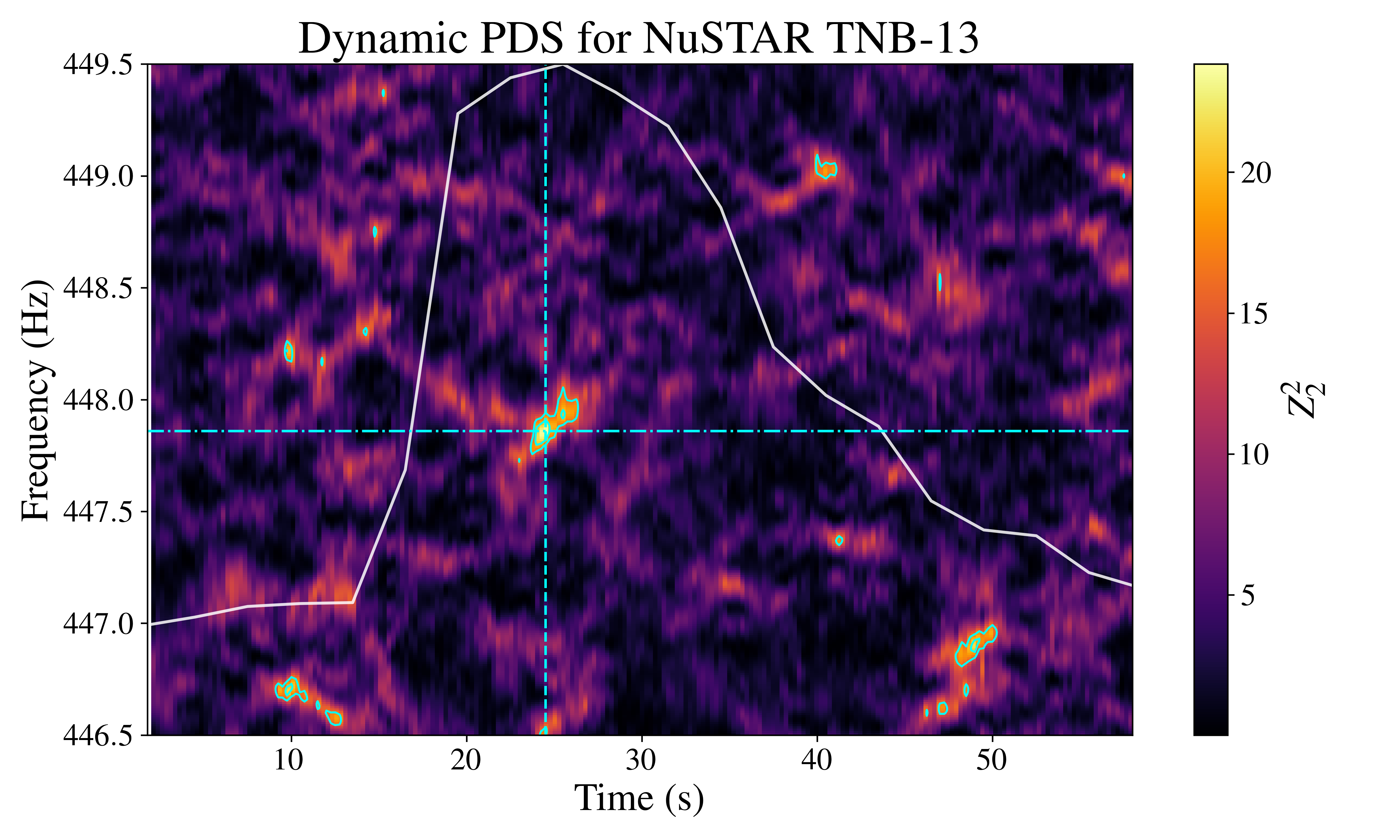}
    \hspace{-0.02\textwidth}
     \includegraphics[width=0.35\columnwidth] {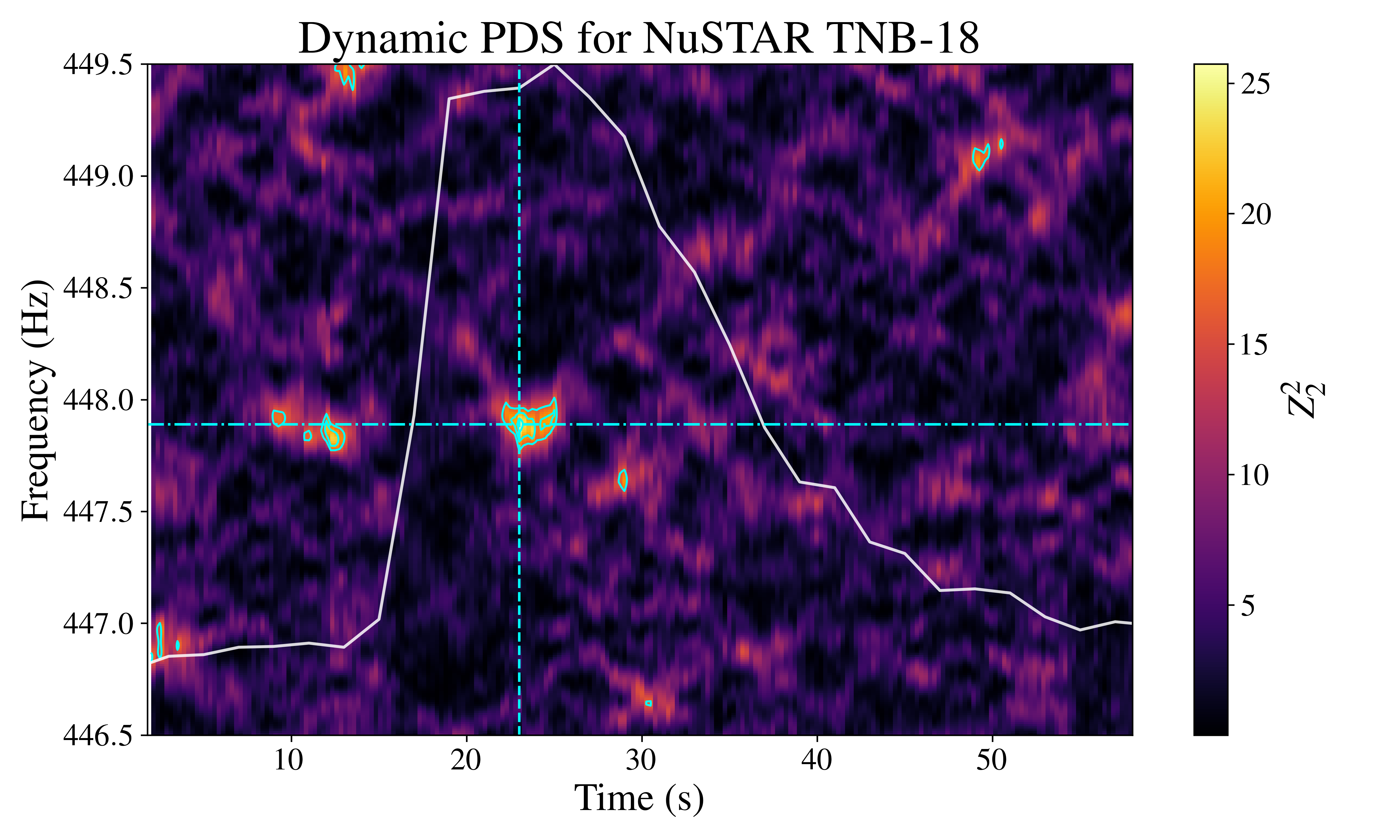}
   { \includegraphics[width=0.33\columnwidth] {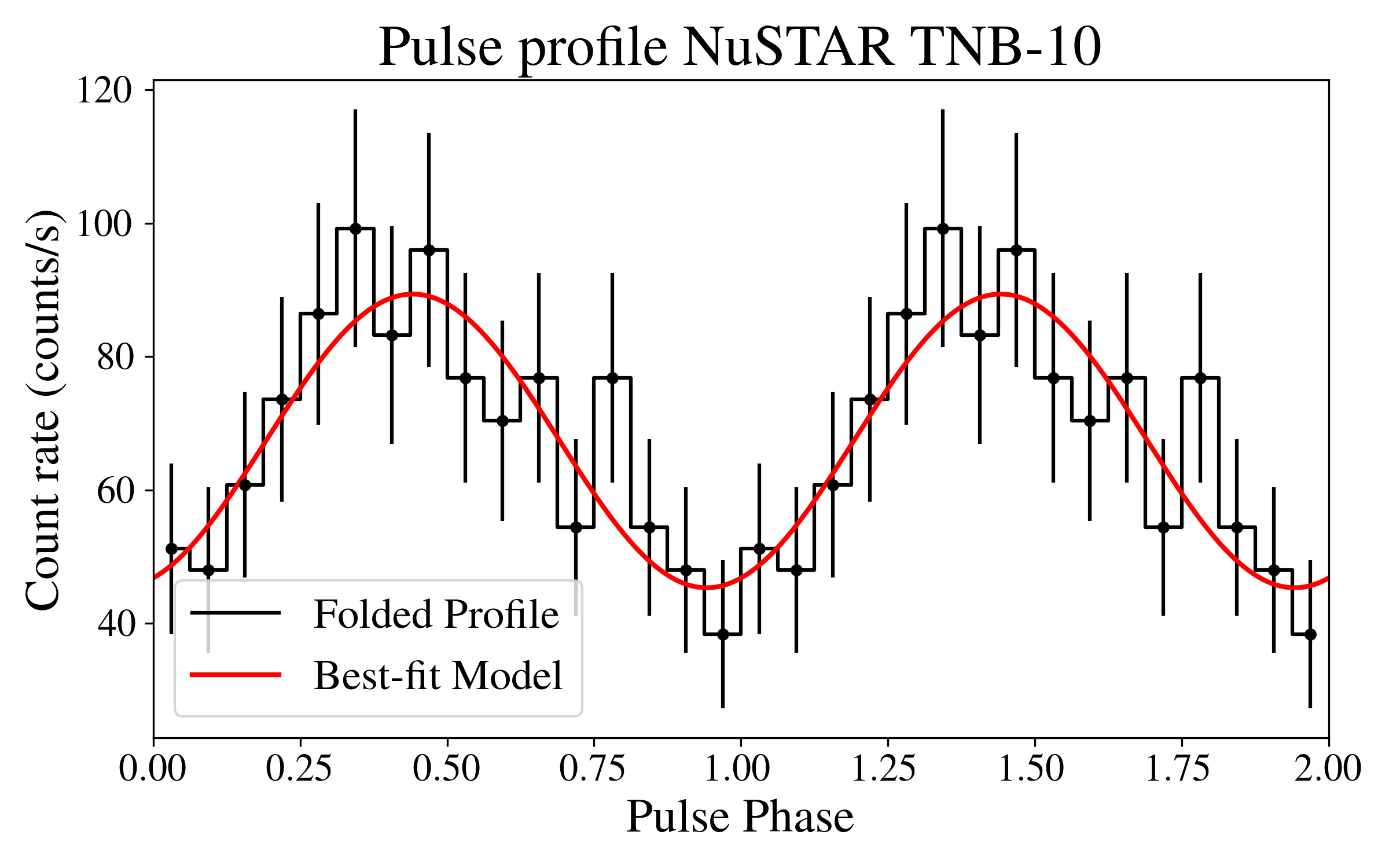}}
  { \includegraphics[width=0.33\columnwidth] {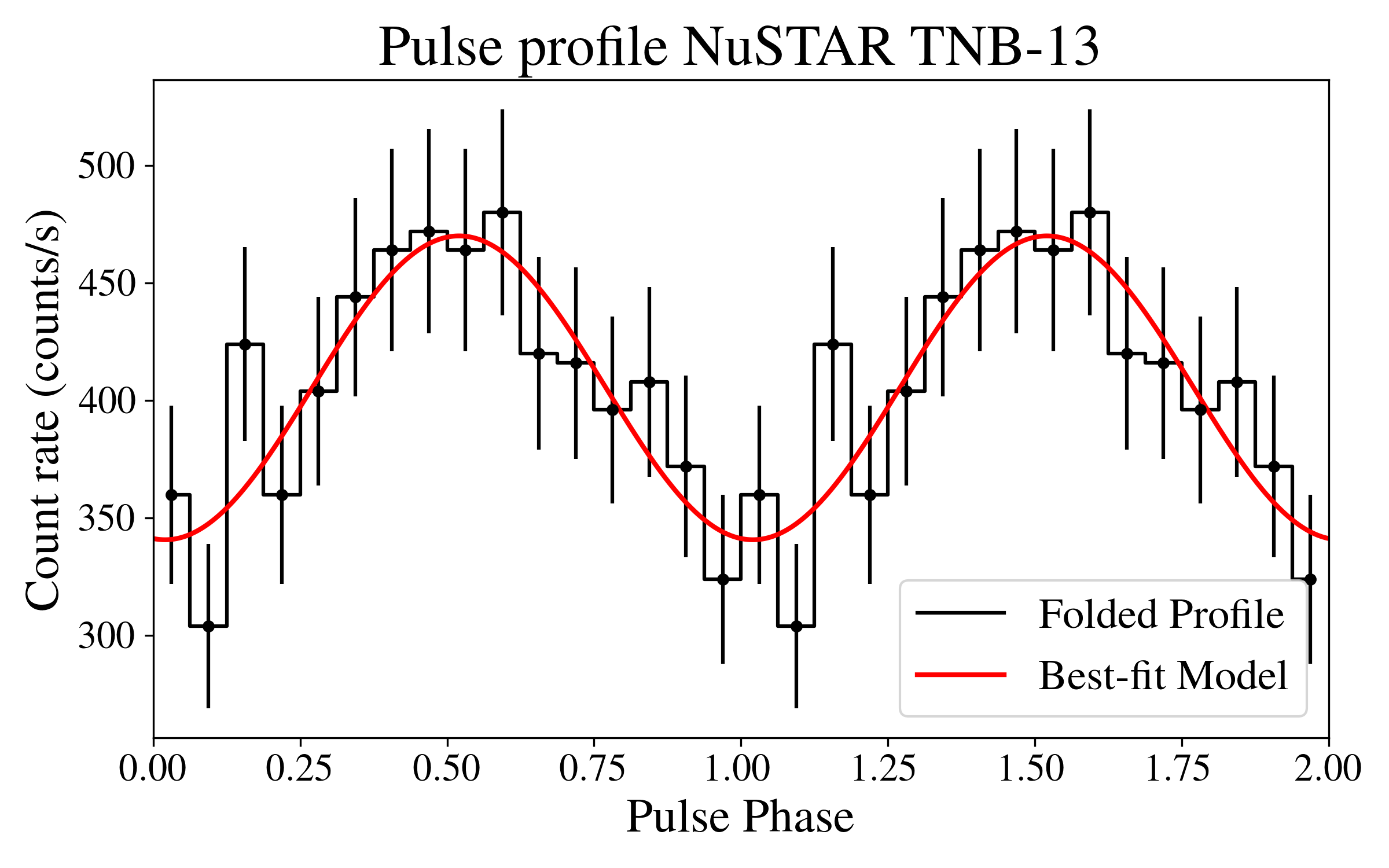}}
  { \includegraphics[width=0.33\columnwidth] {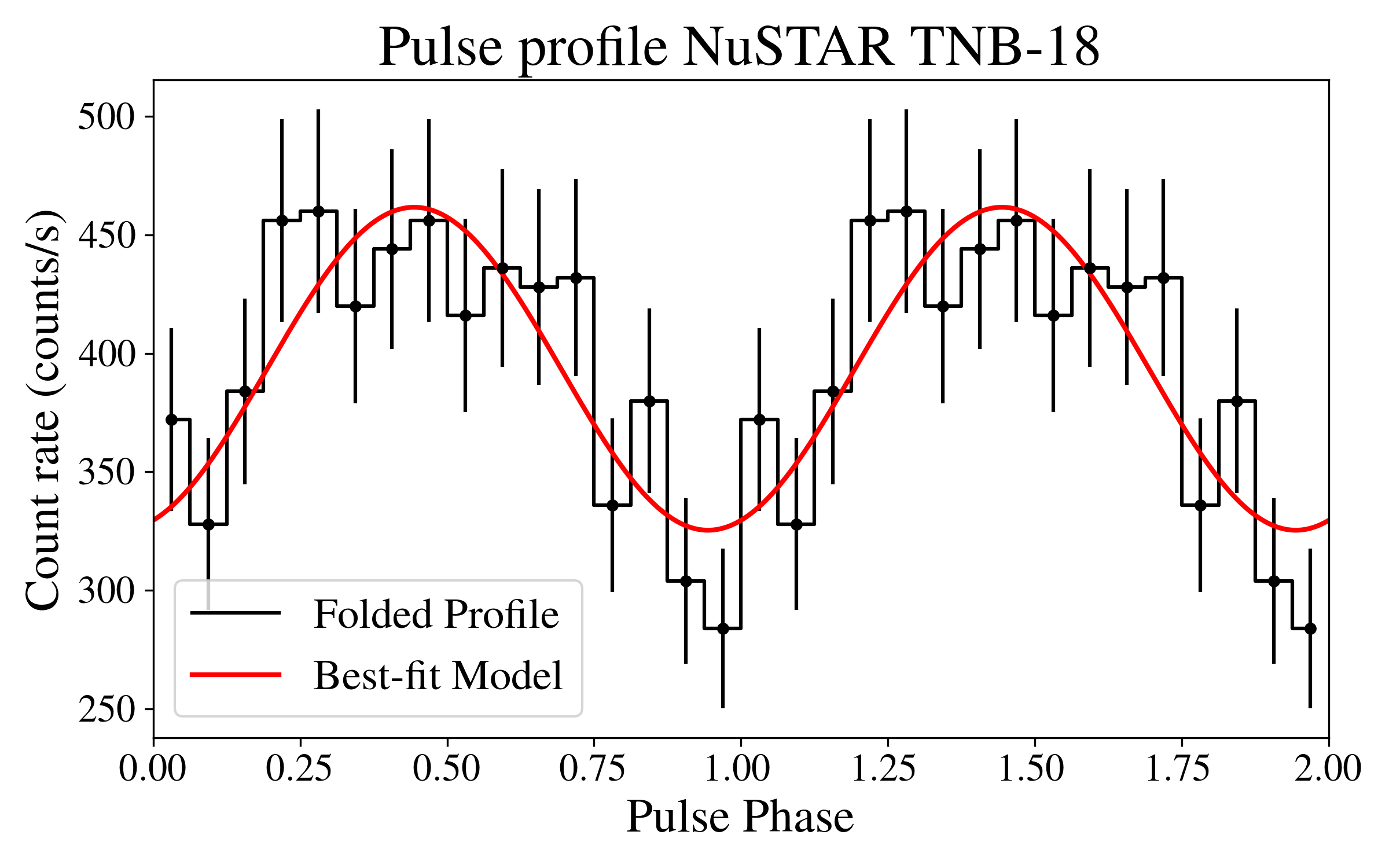}}
 \caption{The dynamic power spectra and the best-fit pulse profiles for the segments (4-6 s duration) with maximum power across six thermonuclear bursts observed with \nustar~ are shown. The dynamic power density spectra of \src{} are overlaid with the corresponding \nustar~ burst light curves. The dynamic PDS is generated from a 4-s sliding window using the $Z^2$ statistic with a step size of 0.25 s over a 3 Hz band centered around the spin frequency. A frequency resolution of 0.01 Hz is employed. Contours are plotted at $Z^2$ levels from 14 to 32 in steps of 4. The horizontal dashed line (cyan) denotes the measured spin frequency, and the vertical dashed line (cyan) denotes the time corresponding to maximum power.}
    \label{fig:DPDS_NUSTAR_all}
\end{figure*}
%%%%%%%%%%%%%%%%%%%%%%%%%%%%%%%%%%%%%%%%%%%
%%%%%%%%%%%%%%%%%%%%%%%%%%%%%%%%%%%%%%%%%%%%
\begin{figure*}
 \includegraphics[width=0.35\columnwidth]{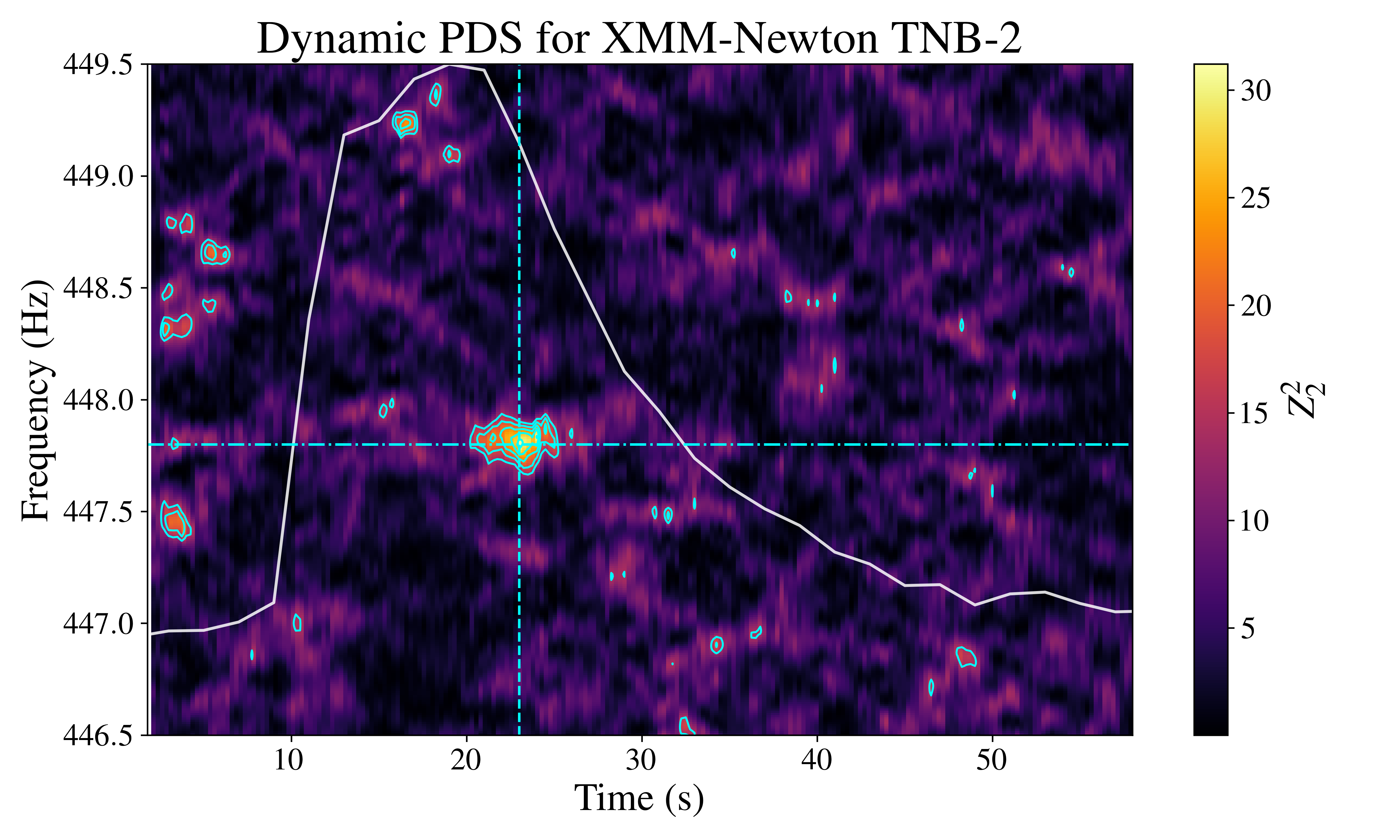}
    \hspace{-0.02\textwidth}
 \includegraphics[width=0.35\columnwidth] {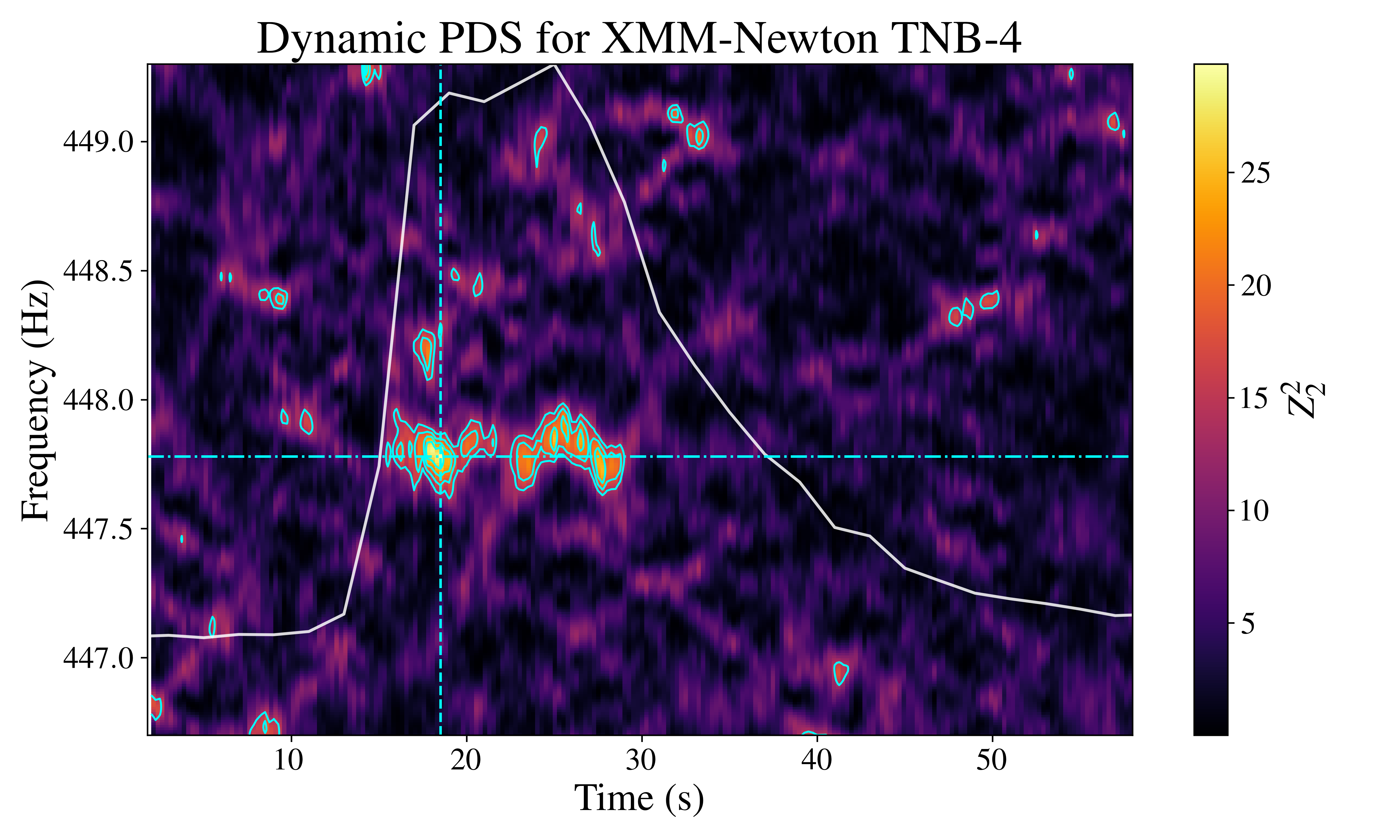}
    \hspace{-0.02\textwidth}
     \includegraphics[width=0.35\columnwidth] {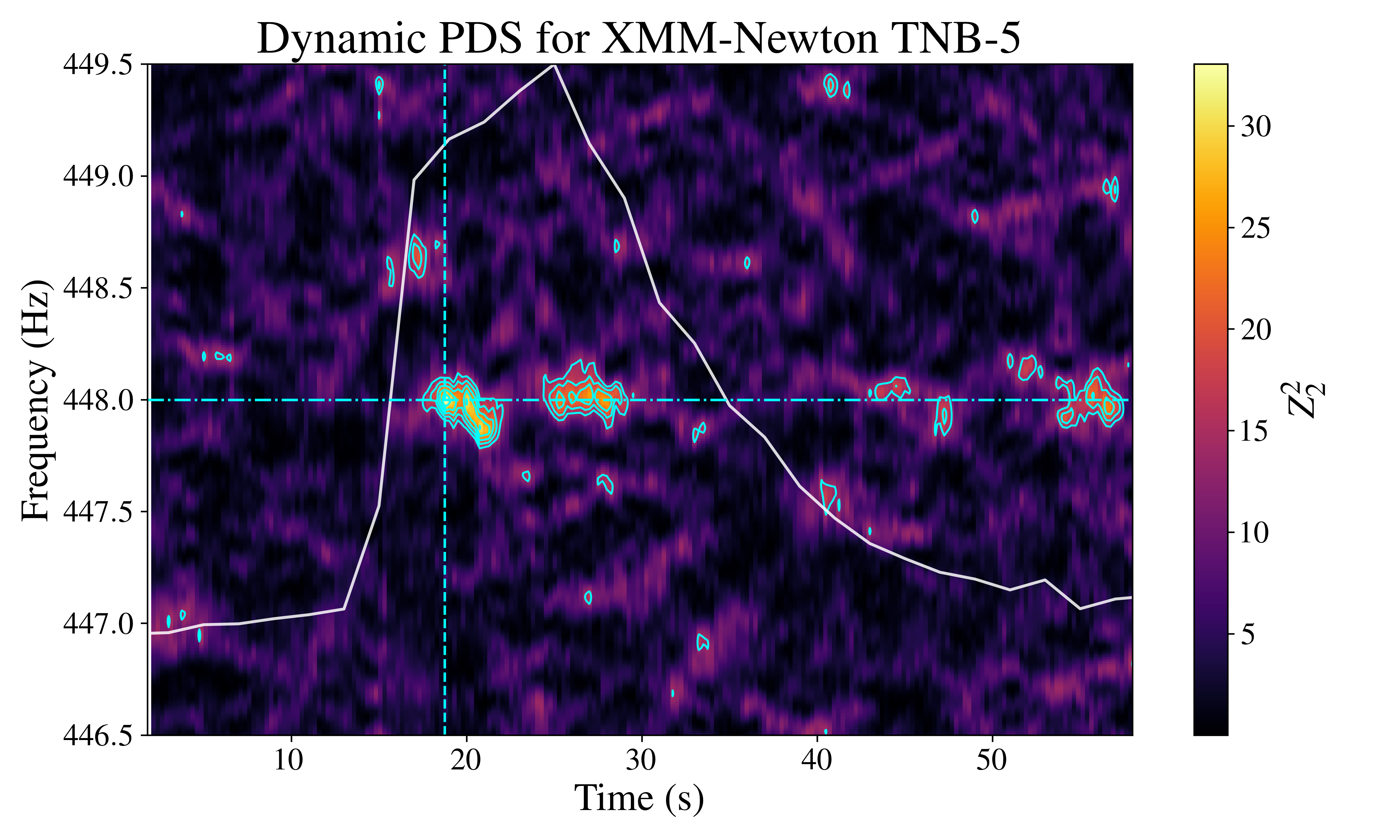}
 \includegraphics[width=0.3\columnwidth] {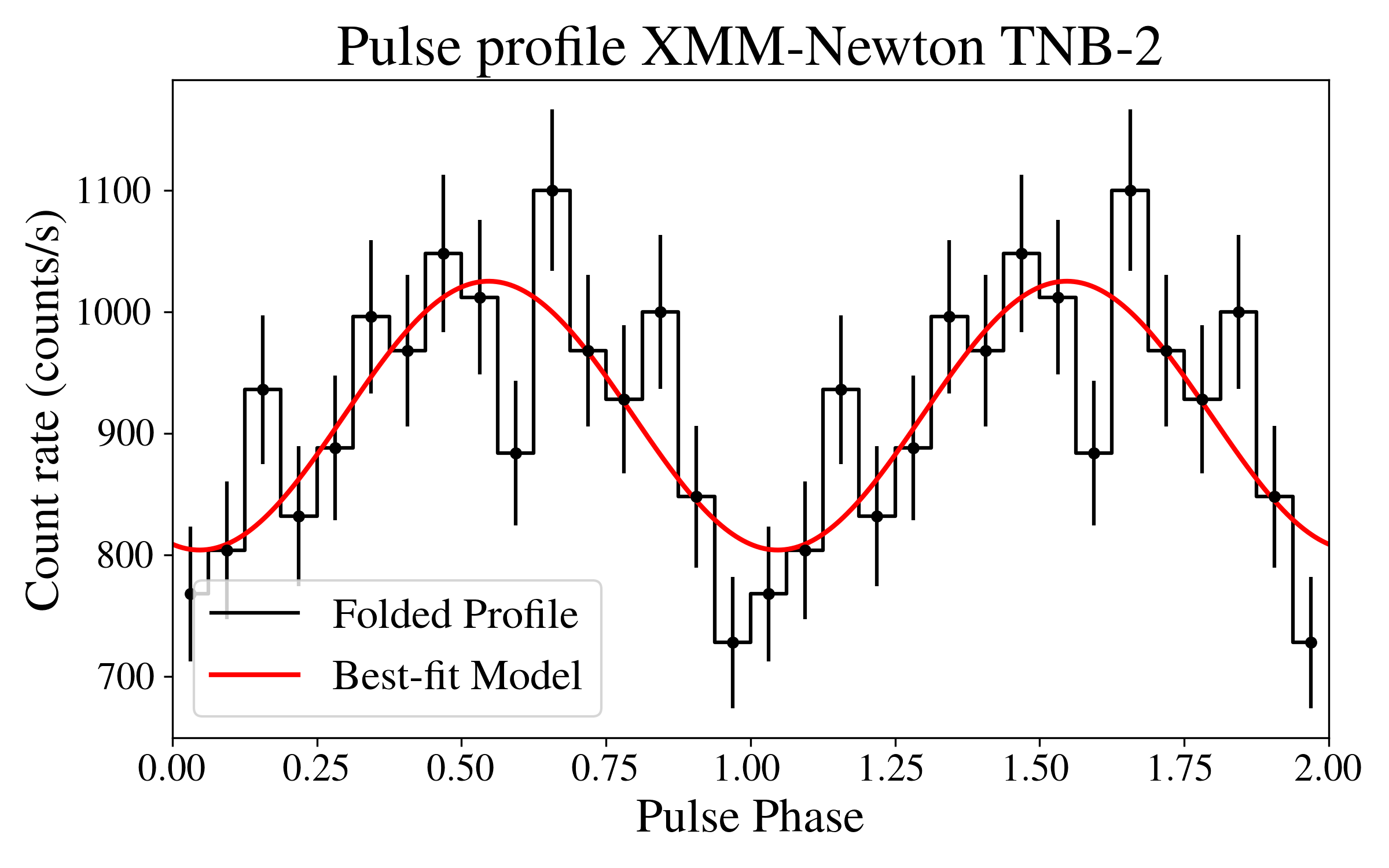}
 \includegraphics[width=0.3\columnwidth] {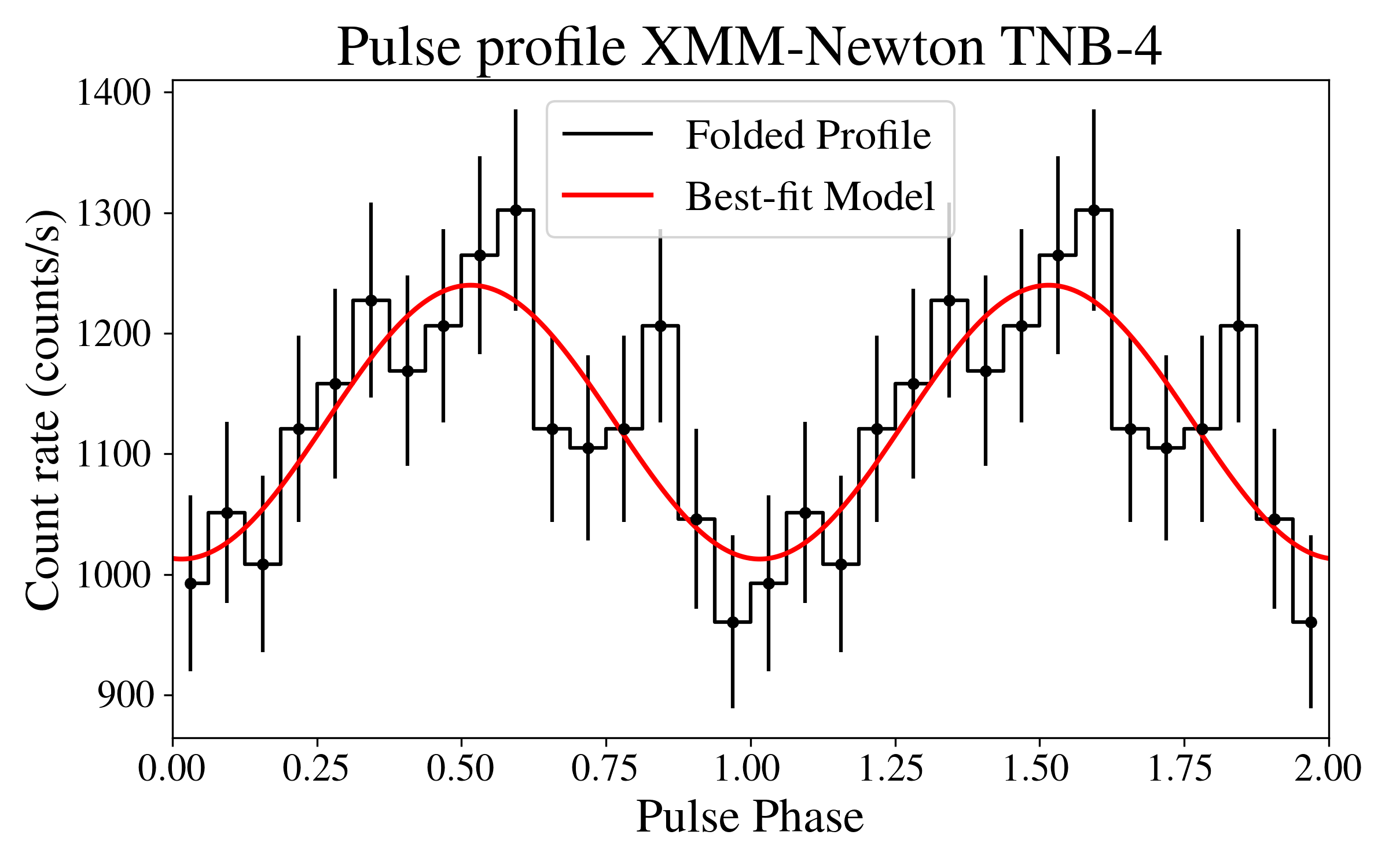}
 \includegraphics[width=0.3\columnwidth] {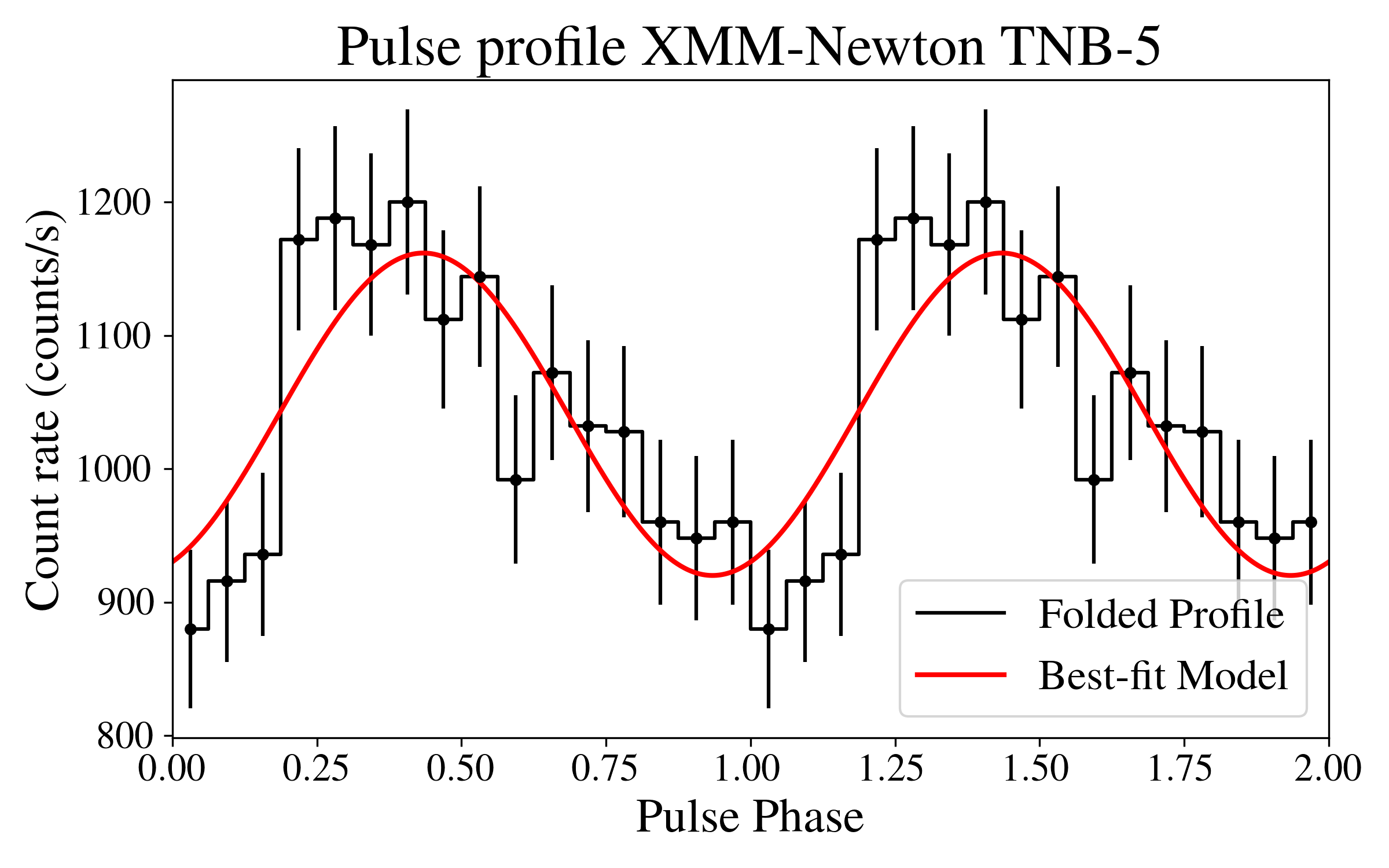}
 \includegraphics[width=0.35\columnwidth] {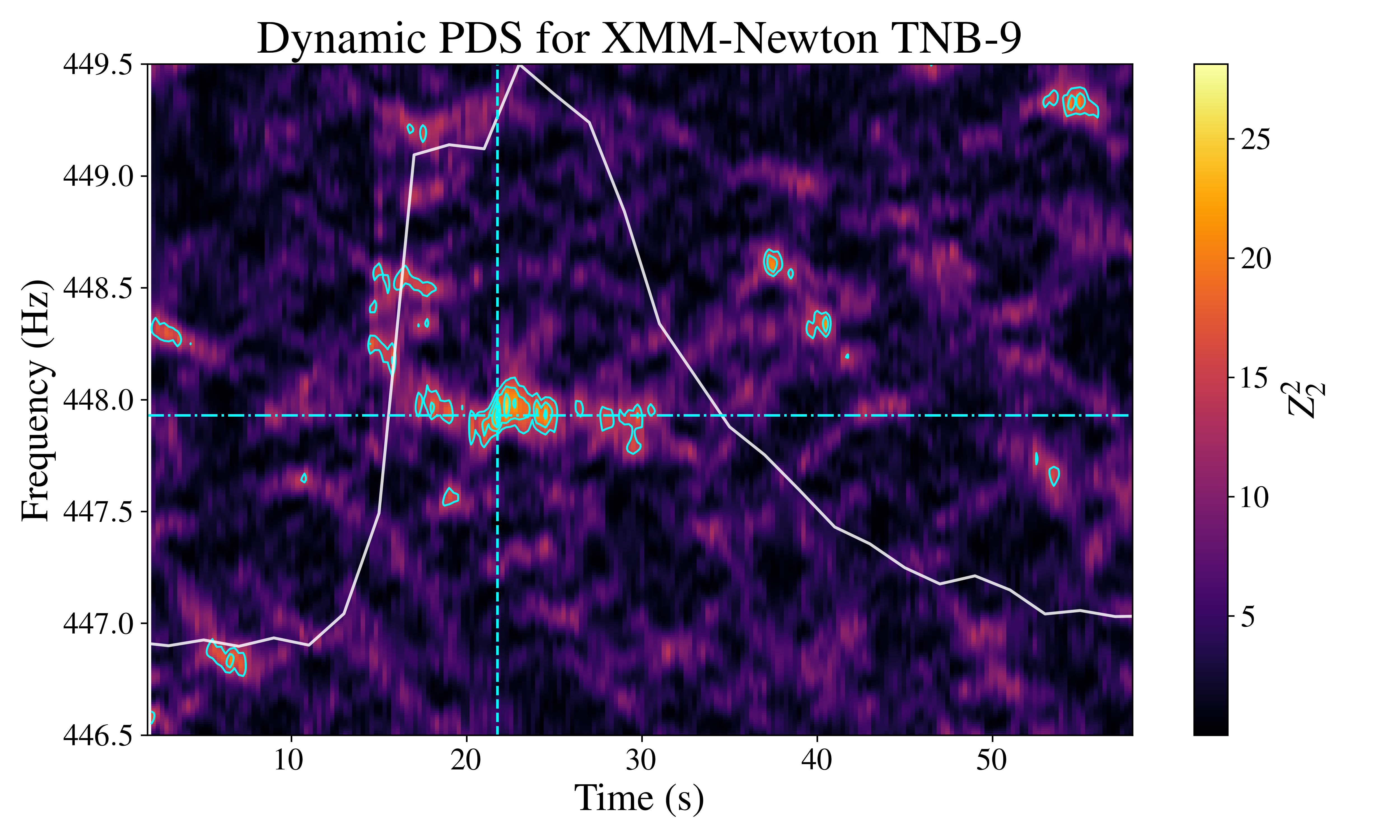}
    \hspace{-0.02\textwidth}
     \includegraphics[width=0.35\columnwidth] {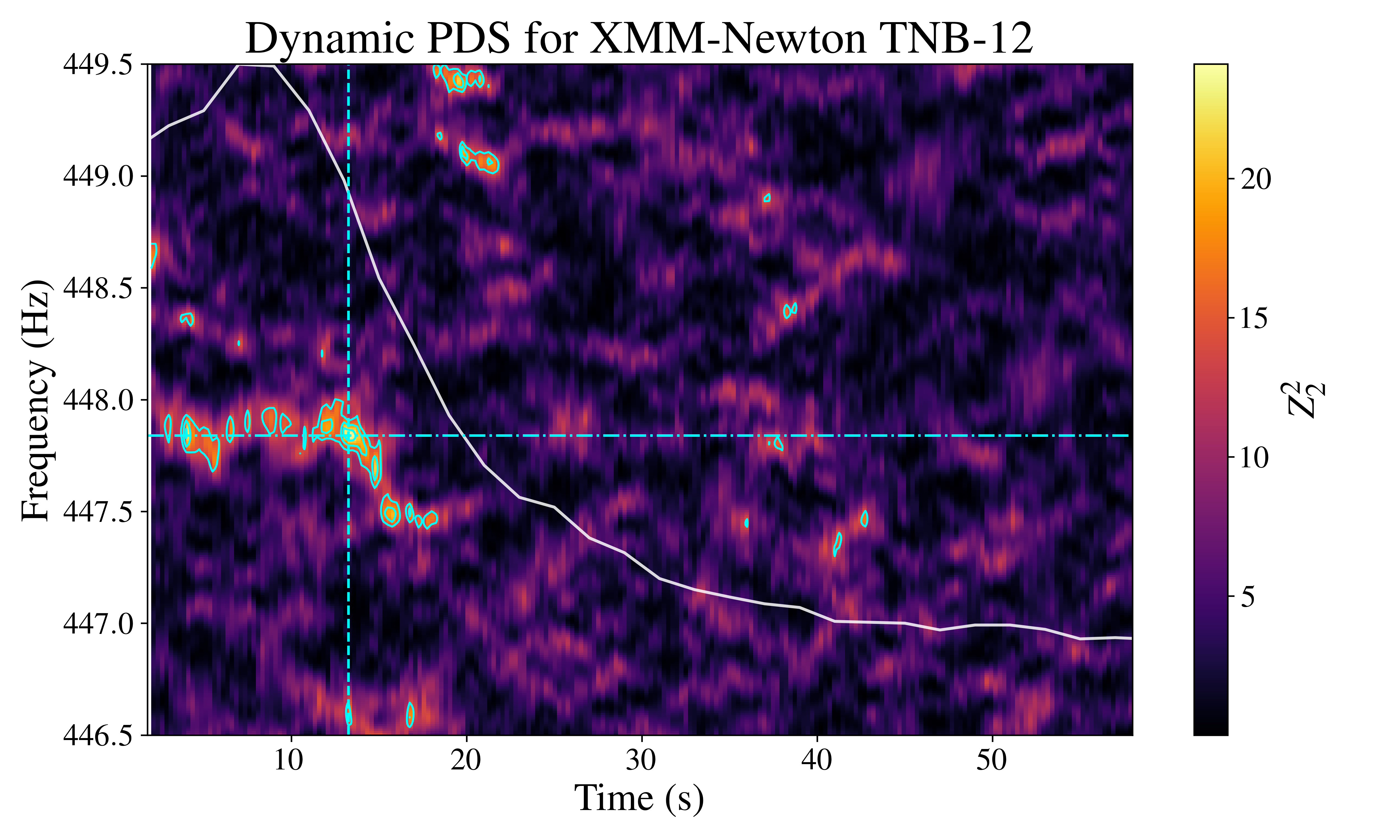}
 \hspace{-0.02\textwidth}
 \includegraphics[width=0.35\columnwidth] {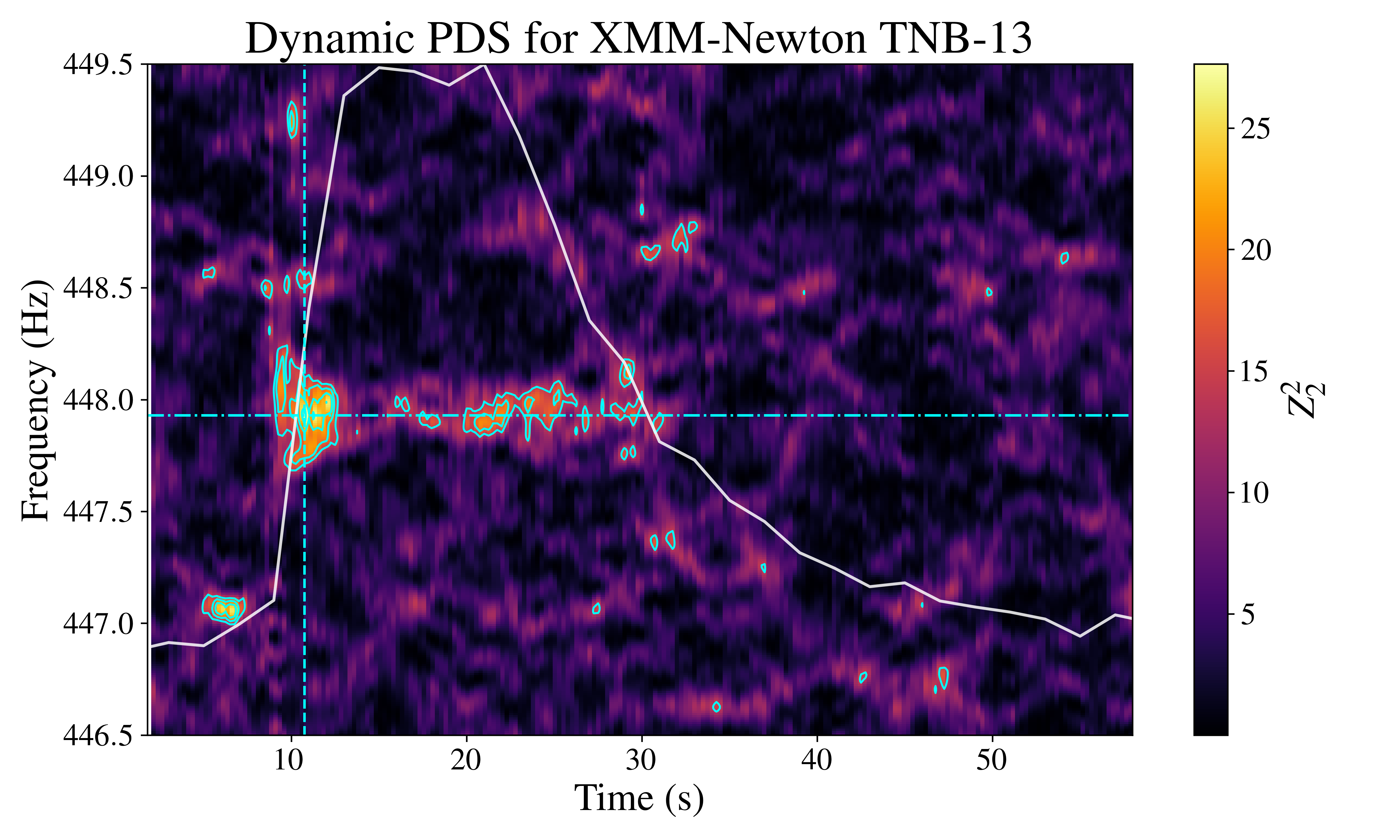}
  \includegraphics[width=0.33\columnwidth] {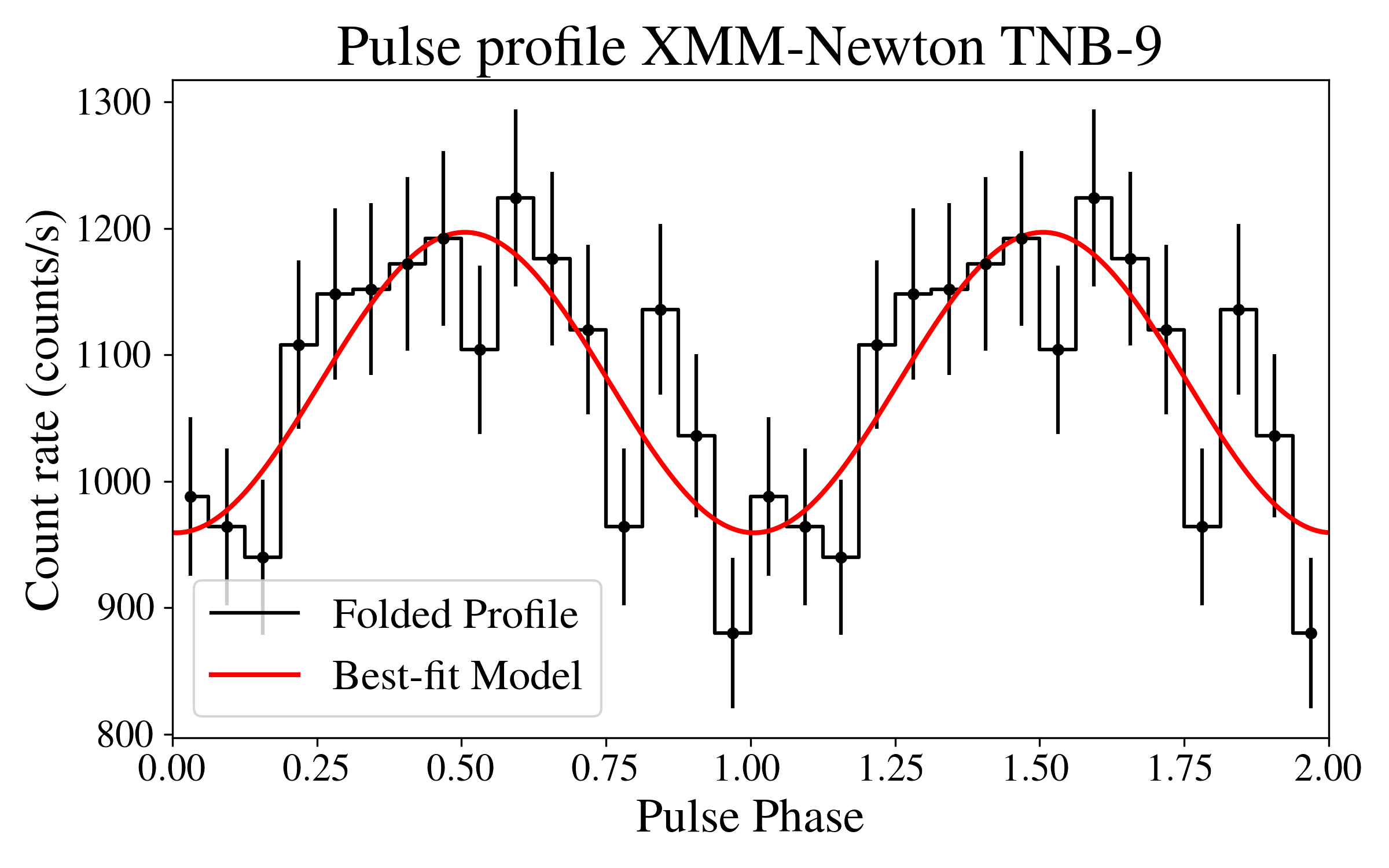}
   \includegraphics[width=0.33\columnwidth] {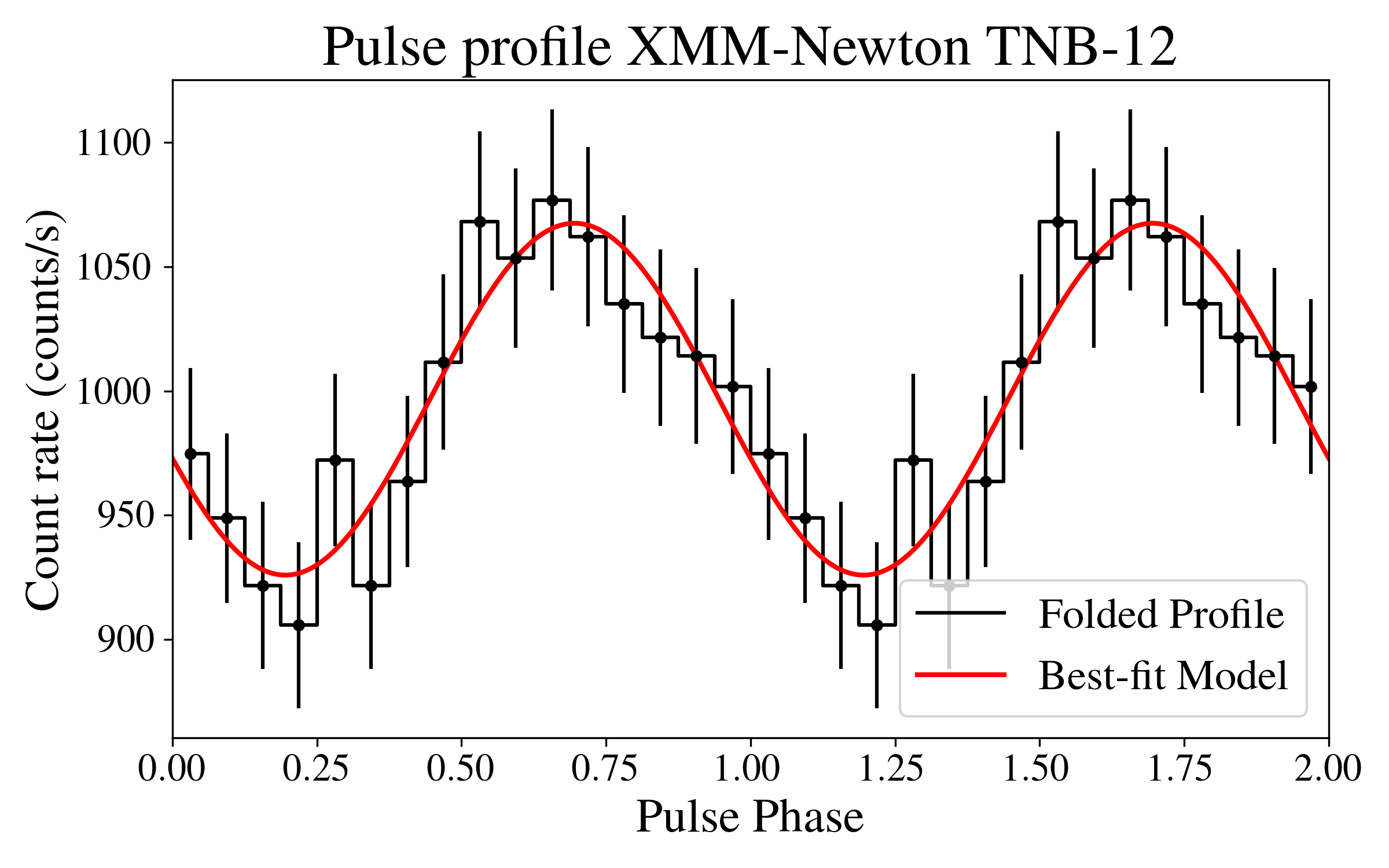}
   \includegraphics[width=0.33\columnwidth]{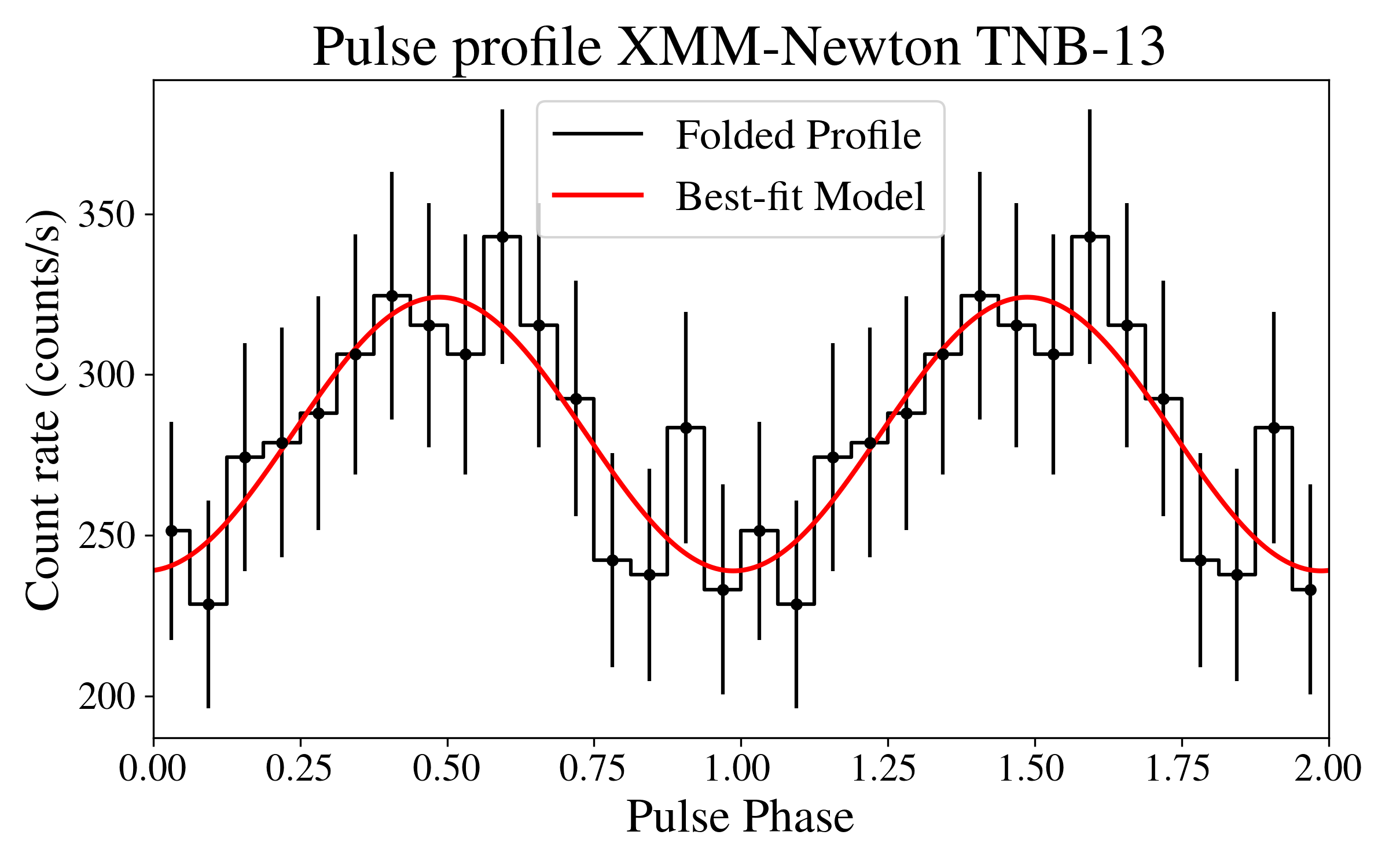}
\caption{The dynamic power density spectra and the best-fit pulse profiles for the segments (4-6 s duration) with maximum power across six thermonuclear bursts observed with \xmm~ are shown. The dynamic power density spectra of \src{} are overlaid with the \xmm~ EPIC-PN burst light curves. The dynamic PDS is generated from a 4-s sliding window using the $Z^2$ statistic with a step size of 0.25 s over a 3 Hz band centered around the spin frequency. A frequency resolution of 0.01 Hz is employed while generating the dynamic PDS. Contours are plotted at $Z^2$ levels from 14 to 32 in steps of 4. The horizontal dashed line (cyan) in the spectra denotes the measured spin frequency, and the vertical dashed line (cyan) denotes the time corresponding to maximum power.}
    \label{fig:DPDS_XMM_all}
\end{figure*}
%%%%%%%%%%%%%%%%%%%%%%%%%%%%%%%%%%%%%%%%%%%%%%%%%
%%%%%%%%%%%%%%%%%%%%%%%%%%%%%%%%%%%%%%%%%%%%%%%%

\bibliography{SRGA_ApJ}{}
\bibliographystyle{aasjournal}

\end{document}